%% file: neurips_2026.tex
\documentclass{article}

\usepackage[numbers, compress, sort]{natbib}
\usepackage[main, final]{neurips_2026}

\usepackage[utf8]{inputenc} 
\usepackage[T1]{fontenc}    
\usepackage{hyperref}       
\usepackage{url}            
\usepackage{booktabs}       
\usepackage{amsfonts}       
\usepackage{nicefrac}       
\usepackage{microtype}      
\usepackage{xcolor}         

\usepackage{algorithm}
\usepackage{algorithmic}

\usepackage{microtype}
\usepackage{graphicx}
\usepackage{subcaption}
\usepackage{booktabs} 
\usepackage{tcolorbox}
\usepackage{xcolor}

\usepackage{hyperref}

\newcommand{\lTilde}{\overset{\scriptsize <}{\sim}}
\newcommand{\gTilde}{\overset{\scriptsize >}{\sim}}
\newcommand{\E}{\mathbb{E}}

\usepackage{amsmath}
\usepackage{amssymb}
\usepackage{mathtools}
\usepackage{amsthm}
\usepackage{tabularx}

\allowdisplaybreaks

\makeatletter
\renewenvironment{proof}[1][\proofname]{\par
  \pushQED{\qed}%
  \normalfont \topsep6\p@\@plus6\p@\relax
  \trivlist
  \item[\hskip\labelsep
        \bfseries 
    #1\@addpunct{.}]\ignorespaces
}{%
  \popQED\endtrivlist\@endpefalse
}
\makeatother

\newcommand{\bigO}{\mathcal{O}}
\definecolor{TUMblue}{RGB}{0,101,189} 
\usepackage{placeins}
\usepackage{enumitem}

\usepackage{thmtools}
\usepackage{thm-restate}

\usepackage{hyperref}
\usepackage[nameinlink,capitalize,noabbrev,]{cleveref}
\crefname{appendix}{Appendix}{Appendices}

\crefname{theorem}{Theorem}{Theorems}
\crefname{proposition}{Proposition}{Propositions}
\crefname{lemma}{Lemma}{Lemmas}
\crefname{corollary}{Corollary}{Corollaries}
\crefname{definition}{Definition}{Definitions}
\crefname{assumption}{Assumption}{Assumptions}
\crefname{remark}{Remark}{Remarks}

\theoremstyle{plain}
\newtheorem{theorem}{Theorem}[section]

\theoremstyle{definition}

\theoremstyle{remark}

\title{The Heavy Hitter Oracle: Enhancing Frequency Estimation in Skewed Data Streams}

\author{%
  Lisa Schmierer \\
  LMU Munich\\
  Munich Center for Machine Learning (MCML)\\ 
  \texttt{lisa.schmierer@lmu.de} \\
  \And
  Ioana-Oriana Bercea \\
  KTH Royal Institute of Technology \\
  Stockholm, Sweden \\
  \texttt{bercea@kth.se} \\
}

\begin{document}

\maketitle

\definecolor{TUMblue}{RGB}{0,101,189} 

\tcbset{
  key_takeaway/.style={
    colback=TUMblue!15,
    colframe=TUMblue,
    boxrule=1pt,
    arc=2mm,
    top=2mm,
    bottom=2mm,
    left=1mm,
    right=1mm,
    width=0.9\columnwidth,
    halign=center,   
    center,           
  }
}

\input{Chapters/00_Abstract}

\input{Chapters/01_Introduction}
\input{Chapters/02_Preliminaries}

\input{Chapters/03_Related_Work}
\input{Chapters/04_Main}
\input{Chapters/05_Experiments}
\input{Chapters/06_Conclusion}

\newpage
\begin{ack}
This work is in part supported by the DAAD programme Konrad Zuse Schools of Excellence in Artificial Intelligence, sponsored by the German Federal Ministry of Research, Technology and Space and the Erasmus+ Programme of the European Union. The second author was supported by the Swedish Research Council (Vetenskapsrådet) under grant no. 2024-05366. We would like to thank the anonymous reviewers for their insightful feedback, as well as Frauke Kreuter and Michael Plainer for their comments.
\end{ack}


\bibliographystyle{abbrvnat}
\bibliography{references}  


\appendix
\crefalias{section}{appendix}
\input{Chapters/appendix_improved_bounds}
\input{Chapters/appendix_large_sketch}

\input{Chapters/appendix_trivial_alternative}
\newpage
\input{Chapters/appendix_experiments}


\newpage
\input{checklist.tex}

\end{document}

%% file: Chapters/00_Abstract.tex
\begin{abstract}
  Estimating element frequencies in data streams is a fundamental problem when processing large amounts of data. Recent approaches suggest learning-augmented algorithms that can
  access a heavy-hitter oracle that knows the most frequent elements. 
  In this paper, we challenge the idea of these complex oracles: First, we show that heavy-hitter predictions do not necessarily yield significant efficiency gains in frequency estimation. We prove that a classical algorithm using slightly more memory, $\Theta(B \log B)$ instead of $\Theta(B)$, matches the performance of learning-augmented methods with perfect predictions. Furthermore, we show that given (perfect) predictions, a trivial mechanism achieves the same performance as state-of-the-art learning-augmented algorithms. On the algorithmic side, we introduce \textit{SpaceR}, a single-pass algorithm that matches the accuracy of existing two-pass prediction-based methods without requiring any predictions. SpaceR uses randomized sampling to identify and track frequent elements in one pass through the data, achieving near-perfect frequency recovery with minimal memory on both synthetic and real-world datasets.
\end{abstract}

%% file: Chapters/01_Introduction.tex
\section{Introduction}
%
Efficiently estimating element frequencies is a common problem in modern large-scale machine learning.
With unlimited resources, the solution to this problem is trivial: Maintain an exact counter for each distinct element. In practice, however, settings comprise far more distinct elements than available memory can track. This fundamental tension --- balancing high-accuracy frequency estimation against strict memory limits --- is the core challenge of scaling these algorithms to massive datasets.

Formally, we are given a stream of $N$ elements from some universe $[n]=\{1,2,\ldots, n\}$. The goal is to estimate the frequency $f_i$ of element $i\in [n]$ using a small amount of space. This fundamental problem of frequency estimation has been widely studied and shown to exhibit surprising trade-offs. For example, one can estimate frequencies within a worst-case error of $\Theta(N/B)$ while using only space $B$~\cite{space_saving_alpha}.  Due to such efficiency benefits, streaming algorithms for frequent items have been successfully used in a variety of large-scale machine learning tasks, such as extreme classification~\cite{amazon}, efficient inference in LLMs~\cite{nlp}, feature hashing and selection~\cite{weinberger2009feature,aghazadeh2018mission}, compression techniques for neural networks, gradient optimizers, and graph-based learning~\cite{talukdar2014scaling, chen2015compressing,spring2019compressing}. They have also been widely used in other areas such as network traffic analysis~\cite{basat2019randomized,holterbach2019blink} and database optimization~\cite{ETH, heddes2024convolution}.

Many applications of interest exhibit strongly skewed distributions, such as the Zipfian distribution \cite{Zipf, Zipf_NLP}. In these streams, a few heavy hitters appear very often, while most elements are rare. By exploiting these underlying characteristics of the data, algorithms can gain further efficiency~\cite{space_saving_alpha, CountSketch}. 

Additionally, a new algorithmic framework has been considered in recent years that uses predictions to go beyond worst-case trade-offs~\cite{mitzenmacher2022algorithms}. These predictions can come from a machine-learned component, domain experts, or historical data. During execution, the main algorithm can access these predictions through an \emph{oracle}. The driving motivation behind such approaches is that realistic data often has an inherent structure that the oracle can detect and can thus be exploited to compute better solutions. This insight has been successfully used for a variety of problems such as low rank approximation~\cite{indyk2021few}, similarity search~\cite{weiss2008spectral, salakhutdinov2009semantic,wang2015learning}, and clustering~\cite{ergun2021learning, nguyen2023improved}. 

In the particular case of frequency estimation,~\citet{ERRORDEF} first introduced learned algorithms by employing an oracle to detect \emph{heavy hitters}, i.e., the top-$k$ most frequent elements in the stream. This allows the system to track exact counts for heavy hitters while processing remaining elements with a classic frequency estimation algorithm. 

\textbf{Realistic oracles.} 
However, the practical and theoretical efficiency of learned predictions for this problem remains unclear. Existing approaches suffer from two fundamental issues:
For one, most prior work does not account for the space required to train and store the oracle. 
This matters especially for streaming algorithms, as they derive their popularity from using far less space than storing the data itself. Building the oracle should therefore not cost significantly more than running the streaming algorithm itself. 

A second concern is with the choice of oracle and the amount of information we expect from it. There are inherent trade-offs between the strength of the oracle and the computational hardness of the problem at hand: the more information an oracle provides, the harder and more expensive it becomes to build. Crucially, constructing the oracle should not be a harder problem than estimating the frequencies itself. Moreover, building the oracle requires multiple passes over the data, which might be infeasible, e.g., in distributed scenarios where the data is stored across several machines.

\textbf{Our contributions.} We study the above trade-offs for heavy-hitter oracles in the context of learned frequency estimation on Zipfian streams~\cite{ERRORDEF,MIT,MIT_prime}.
Our main contributions are:

\begin{itemize}
    
    \item We prove that a classical single-pass algorithm can recover the advantage of learning with only a logarithmic increase in space, achieving the same error with space $\Theta(B \log B)$ as learned methods do with $\Theta(B)$.

    \item We show both theoretically and experimentally that it is possible to match other state-of-the-art (SOTA) learned algorithms with a much simpler approach, which we refer to as \textit{Trivial Alternative}.

    \item We introduce \textit{SpaceR}, a new oracle-free streaming algorithm inspired by heavy-hitter methods, that experimentally matches the performance of all known learned algorithms. 
\end{itemize}

\textbf{Paper outline.} The paper proceeds as follows: after establishing notation and reviewing streaming baselines in~\cref{Preliminaries}, we review related work in~\cref{Related_Work}. ~\cref{theory1} discusses an improved analysis of the unlearned SOTA CountSketch++ algorithm.  In ~\cref{theory2} we then analyze a trivial alternative to refined learning-augmented algorithms.~\cref{SpaceR} introduces SpaceR, a novel approach to frequency estimation in one pass. Finally, we evaluate the presented methods in~\cref{experiments}, before~\cref{conclusion} concludes the paper.

%% file: Chapters/02_Preliminaries.tex
\section{Preliminaries}\label{Preliminaries}

\textbf{The frequency estimation problem.} The stream consists of a sequence of $N$ elements $x_1,\ldots, x_N \in [n]=\{1,\ldots, n\}$. Let $f_i$ denote the number of times item $i$ appears in the stream. For simplicity, we assume without loss of generality that $f_1 \geq f_2 \geq f_3 \geq \cdots \geq f_n$.
The streaming algorithm processes the elements of the stream one by one. At the end of the stream, it answers queries of the form ``What is the frequency of $i$?''. We let $\hat{f}_i$ denote the estimate it returns. In the setting with predictions, the algorithm has additional access to a heavy-hitter oracle that receives a query element $i$ and returns ``yes/no'' depending on whether it believes element $i$ is a heavy-hitter element or not. In all practical implementations, we note that these predictions are represented as an array of elements that the algorithm receives before it sees the stream.

\textbf{Notation.} Throughout our proofs, we will employ two main parameters: $B$ will tune the space budget we use for the streaming algorithm, and $k$ will denote the number of predictions we require from the oracle, i.e., a perfect oracle would return the identity of the top-$k$ most frequent items. We note that the space complexity of our algorithms will be counted in words of memory.

\textbf{Estimation error.} We employ the weighted error model from previous work \cite{ERRORDEF,MIT,MIT_prime}
\[\textit{Weighted Error} = \frac{1}{N} \sum_{i\in [n]} f_i \cdot |\hat{f}_i - f_i|\]
This corresponds to the average error over all elements when the query probability is proportional to the original frequency of the element in the stream.

\textbf{Zipfian distribution.} 
In a Zipfian distribution, the frequency of an element is inversely proportional to its rank in the frequency table, i.e., the $i^{th}$ largest frequency $f_i$ is proportional to $A/i^\alpha$, for some $A$ and $\alpha$, where $A$ is a scaling factor and $\alpha$ controls the skewness of the distribution. Following previous work, we consider the case $\alpha = 1$ and obtain $A = N/H_n \approx N/\log n$. 
Similarly, throughout our theoretical analysis, we assume for simplicity that $f_i = 1/i \cdot N/\log n$.

\textbf{CountSketch.}\label{CountSketch} 
The hash-based CountSketch (CS) algorithm~\cite{CountSketch} allocates $d$ arrays of $w$ counters each, for a total space of $B=d\cdot w$ counters. It then sets up $d$ independent hash functions $h_1, \ldots, h_d : [n] \to [w]$ and $d$ independent sign hash functions $s_1, \ldots, s_d : [n] \to \{-1, +1\}$. We denote the $d \times w$ table by $C$ and define each counter in the table as $$C[\ell, b] = \sum_{i \in [n]} [h_\ell(i) = b] \cdot s_\ell(i) \cdot f_i \;.$$ 
In other words, as we process element $i$ in the stream, we increase each of its $d$ random counters by the value of the corresponding sign function $s_\ell(i)$. When querying for the frequency of $i$, the algorithm returns an estimate  $$\tilde{f}_i = \text{median}_{\ell \in [d]} s_\ell(i) \cdot C[\ell, h_\ell(i)] \;.$$

\textbf{SpaceSaving.} The second algorithm of relevance to us is the SpaceSaving algorithm of~\citet{SpaceSaving}. It is a deterministic algorithm that maintains a heap of $m$ elements, each associated with a counter. When it processes element $i$ in the stream, it first checks if the element is already tracked in the heap. If so, it increases its corresponding counter. Otherwise, it swaps $i$ with the element with the minimum counter and increases the counter by one. 

\textbf{Learning-augmented algorithms.} Learned or learning-augmented algorithms are a subgroup of algorithms that assume access to an oracle that identifies the heavy hitters in a stream. 
Upon arrival of an element, the algorithm queries the oracle: if the element is among the predicted heavy hitters, it is tracked with exact counters, while all other elements are handled by the default streaming algorithm.
To transform a classical algorithm into its learned counterpart, the system splits the memory: one half maintains a heap to track exact counts for predicted heavy hitters, while the remaining space initializes the base algorithm (e.g., CountSketch) to process all other elements. Consequently, predicted heavy hitters bypass the base algorithm entirely during queries and updates.

\textbf{Learned and unlearned CountSketch++.} \label{CS++_old} The unlearned CountSketch++ (CS++) algorithm~\cite{MIT} employs $\log \log n$ small CountSketches of size $3 \times \frac{B}{6 \log \log n}$ and one big CountSketch of size $3 \times\frac{B}{6}$ for a total space of $B$ words. As it processes elements, it inserts them into all tables. Upon a query, it first checks the small tables and takes the median $\tilde{f}_i$ over all returned estimates. If this median $\tilde{f}_i$ is smaller than some threshold $\tau = \bigO((\log\log n)/B)$, CS++ outputs zero as the final estimate for the element's frequency. Intuitively, such elements should have small frequencies, and returning $0$ on them therefore lowers the expected error. If  $\tilde{f}_i$ is greater than $\tau$, then CS++ returns the estimate from the big CS table, which has better error guarantees than the smaller tables.

The learned CS++ version allocates half of the space budget to track exact frequencies for $\Theta(B)$ predicted heavy hitters. With the remaining space, it initializes a classic CS++ sketch to process all other elements.

%% file: Chapters/03_Related_Work.tex
\section{Related work}\label{Related_Work}
The frequency estimation problem has been widely studied as a fundamental streaming problem without predictions~\cite{manku2002approximate, demaine2002frequency, karp2003simple, space_saving_alpha, Braverman2016BPTreeA, anderson2017high} (see also~\citet{Overview} for an overview). 
Most widely implemented in practice are four algorithms~\cite{zaharia2016apache, medjedovic2022algorithms,datasketches}. Out of the four, the most relevant algorithms for this work are the hash-based CountSketch (CS) algorithm~\cite{CountSketch} and the deterministic SpaceSaving algorithm~\cite{SpaceSaving}. Other notable algorithms include CountMin~\cite{CountMin} and Misra-Gries~\cite{MisraGries}. The CountMin algorithm is hash-based and closest to CS in nature. As CountMin theoretically and empirically performs worse than CS on Zipfian data in both the learned and classical versions \cite{MIT}, we do not discuss it further in this work. The Misra-Gries algorithm is a deterministic algorithm originally developed for detecting heavy hitters (similar to SpaceSaving). We evaluate its efficiency as a possible oracle, but find that it performs weaker than other oracles. 

\textbf{The learned approach.}
In the particular case of frequency estimation,~\citet{ERRORDEF} were the first to consider this learned approach. The oracle they employ detects \emph{heavy-hitter} elements, i.e., the top-$k$ most frequent elements in the stream. They showed that this leads to more efficient estimates than SOTA when the input data follows a Zipfian distribution. In their work,~\citet{ERRORDEF} study the benefits of using CountMin with the oracle, while~\citet{du2021putting} study how to best optimize the oracle in order to achieve the guarantees of~\citet{ERRORDEF}. 

Subsequent work by~\citet{MIT} tightens the theoretical bounds on CountMin from~\citet{ERRORDEF} and further gives tight bounds for learned and classic CS on Zipfian data. They show that, under the same Zipfian assumption, one can essentially reduce the worst-case error by a multiplicative factor of $(\log n)^2$ with a streaming algorithm which asymptotically uses the same amount of space as its unlearned counterpart (see \cref{tab:weighted-error-comparison}). Furthermore,~\citet{MIT} find that CS outperforms CountMin (in both learned and classic settings), but Learned CS has an advantage over CS only in the high-accuracy regime, when the space $B$ of the algorithm is some polynomial fraction of $n$, i.e. $B=n^{1-o(1)}$. To address the more prevalent regime of $B=\textit{poly}(\log n)$, they introduce CountSketch++ (CS++), which they show, theoretically and empirically, outperforms both CountMin and CS in both learned and classic settings.
\begin{table*}
  \caption{Comparison of expected weighted error bounds for $B$ words of memory and Zipfian frequencies, i.e., $f_i \propto \frac{1}{i}$.}
  \label{tab:weighted-error-comparison}
  \begin{centering}
    \begin{small}
    \resizebox{\textwidth}{!}{%
      \begin{sc}
        \begin{tabular}{lcccr}
          \toprule
          Algorithm  & Weighted Error        & Use Predictions?      & Reference  \\
          \midrule
          CountSketch & $\Theta\left(\frac{N}{\log(n)}\cdot \frac{1}{B}\right)$ & No & \cite{MIT} \\
          Learned CountSketch & $\Theta\left(\frac{N}{\log^2(n)}\cdot \frac{\log(n/B)}{B }\right)$ & Yes & \cite{MIT} \\ \midrule
          CountSketch++ & $\mathcal{O}\left(\frac{N}{\log^2(n)}\cdot \frac{\log B + \mathrm{poly}(\log \log n)}{B}\right)$ & No & \cite{MIT} \\
          CountSketch++ \textit{(Improved Bound)} & $\mathcal{O}\left(\frac{N}{\log^2(n)}\cdot \frac{\log B}{B }\right)$ & No & \cref{thm:improved_error}\\
          Learned CountSketch++ & $\mathcal{O}\left(\frac{N}{\log^2(n)}\cdot \frac{1}{B}\right)$ & Yes & \cite{MIT}\\
          Trivial Alternative & $\mathcal{O}\left(\frac{N}{\log^2(n)}\cdot \frac{1}{B}\right)$ & Yes & \cref{thm:triv_alternative} \\
          \bottomrule
        \end{tabular}
      \end{sc}
      }
    \end{small}
  \end{centering}
\end{table*}
Recently,~\citet{MIT_prime} further considered a learned version of Misra-Gries and generalized it to the problem of finding frequent directions. 
They show that Learned Misra-Gries matches the performance of CS++ theoretically and empirically, while having the advantage of being deterministic and simpler in design. Finally,~\citet{shahout2024learning} present a learned variant of SpaceSaving, but their trade-offs are not comparable to those of the previous works.

Existing learned frameworks \cite{MIT, ERRORDEF, MIT_prime} often rely on neural networks as oracles, which introduce significant overhead. First, their memory requirements often conflict with the limited local storage inherent to streaming settings. This might make neural networks inappropriate for applications where one cannot afford to store the oracle alongside the algorithm's local memory. Second, even when the neural net can be stored, e.g., in external memory, neural networks require domain-specific training. 
This process is resource-intensive and needs substantial training data to produce a reliable predictor. We overcome these limitations by evaluating lightweight oracles that fit within the algorithm’s own space budget and do not require exhaustive training for different domains.

A second limitation of the framework proposed by \citet{ERRORDEF} is that it requires multiple passes over the data to construct the oracle. Such multi-pass requirements may not be feasible in practice, e.g., in distributed scenarios where data is partitioned across several machines. We address this by investigating one-pass alternatives that maintain high accuracy without the need for re-scanning the data.

%% file: Chapters/04_Main.tex
\section{Improved analysis on CS++ without predictions} \label{theory1}


To address the issues motivated above, we first aim to understand the extent of the advantage when using the learned version versus the unlearned version. We begin by re-examining the CS++ algorithm, a SOTA approach for Zipfian data.
The original analysis in \citet{MIT} shows that CS++ without predictions achieves an expected weighted error of $\mathcal{O}\left(\frac{N}{\log^2(n)}\cdot \frac{\log B + \mathrm{poly}(\log \log n)}{B}\right)$ on Zipfian streams whenever the space used satisfies $B \geq \log n$.  We show that this upper bound on CS++ without predictions can be  further improved as such:

\begin{restatable}{theorem}{theoImprov} \label{thm:improved_error}
   CountSketch++ with space $B \geq \log n$ updated over a Zipfian stream, has expected weighted error: \[\mathop{\mathbb{E}}[\textit{Weighted Error}] = \bigO \Bigg(\frac{N}{\log^2(n)}\cdot \frac{\log B }{B} \Bigg) \;.\]
\end{restatable}

The proof of \cref{thm:improved_error} can be found in~\cref{appendix:improved_analysis}. The improvement over the previous bound is most significant in the regime in which $B=\textit{poly}(\log n)$, since it avoids the influence of the exponents in the $\textit{poly}(\log\log n)$ term in~\cite{MIT}. Our analysis shows that unlearned CS++ already asymptotically outperforms \textit{learned} CS (with perfect predictions) for any $B$ that is sublinear in $n$. We note that~\citet{MIT} also give matching lower bounds for the expected error of learned CS.~\cref{tab:weighted-error-comparison} gives an overview of the expected weighted error bounds for the algorithms discussed in this work. 

As a corollary of the above, we get that, if we allocate slightly more space to the classic CS++ (without predictions), we get the same bound on the expected weighted error as if we employed perfect predictions:

\begin{restatable}{corollary}{corLearnedSketch} \label{cor}
CountSketch++ (without predictions) with a space budget of $\bigO(B \log B)$ on a Zipfian data stream achieves the same weighted error as its learned counterpart with perfect predictions and space budget $B$, that is:
\[\mathop{\mathbb{E}}[\textit{Weighted Error}] = \bigO \left( \frac{N}{\log^2(n)} \cdot \frac{1}{B} \right) \;.\]
\end{restatable}

We note here that the main difference between the classic and learned versions of CS++ lies in the fact that, in the learned version, the predicted heavy hitters are never processed by the unlearned CS++. Because of this, the learned CS++ version can more accurately filter out the rare elements (i.e., whose estimated frequency $\tilde{f}_i$ from the small tables is smaller than $\tau$). This is due to fewer collisions with heavy elements in the small tables. Nevertheless, such mistakes contribute very little to the overall weighted error, since the elements have small frequencies to begin with. We believe this is why we are able to obtain comparable gains at such a low increase in space.

We find this trade-off to be somewhat surprising because we are not aware of any algorithms that are guaranteed to find the top-$(B/2)$ heavy hitters in space $o(B\log B)$, which the learned CS++ sketch would have to do if it were to compete with the unlearned version in its overall space usage.

\section{A trivial alternative to learned CS++}\label{theory2}
Corollary~\ref{cor} suggests that most of the weighted error should come from elements of large frequencies for which we do not have good estimates. 
We show that this is indeed the case, in a rather extreme way. Specifically, we consider an algorithm that takes the identity of the top-$(B/2)$ elements (i.e., perfect predictions), keeps count of their frequencies exactly, and then outputs zero for the remaining elements. We refer to this mechanism as the \textit{Trivial Alternative} in the following and show that it asymptotically matches the performance of a similarly-sized learned CS++ with perfect predictions:

\begin{theorem}\label{thm:triv_alternative}
    Consider the Trivial Alternative that keeps exact counts for the top-$(B/2)$ elements predicted by an oracle, updated over Zipfian data. Then:
    \[\mathop{\mathbb{E}}[\textit{Weighted Error}]= \bigO \left( \frac{N}{\log^2(n)}\cdot \frac{1}{B} \right) \;. \]
\end{theorem}
The proof can be found in~\cref{appendix:trivial_alternative}. This theoretical insight is supported by our experimental results in~\Cref{sec:experiment1}. Furthermore, observations on dynamic re-sizing of counters for frequency estimation without predictions show that even without predictions, there might exist settings in which it is possible to reduce the error by returning 0 as an estimate for all elements~\cite{NoMeasurementsAppendix}. 

\FloatBarrier

\section{SpaceR}\label{SpaceR}
The findings above naturally lead to the following question: if such a simple mechanism can match the performance of more complex learning-augmented algorithms, is a heavy-hitter oracle truly necessary? Or can a single pass provide enough information to match the performance of SOTA learned algorithms?
To address this question, we introduce \textit{SpaceR} and show that, experimentally, it matches SOTA both when used as an oracle for predictions and when used alone as an oracle-free algorithm.

\begin{figure}
        \centering
        \includegraphics[width=\textwidth]{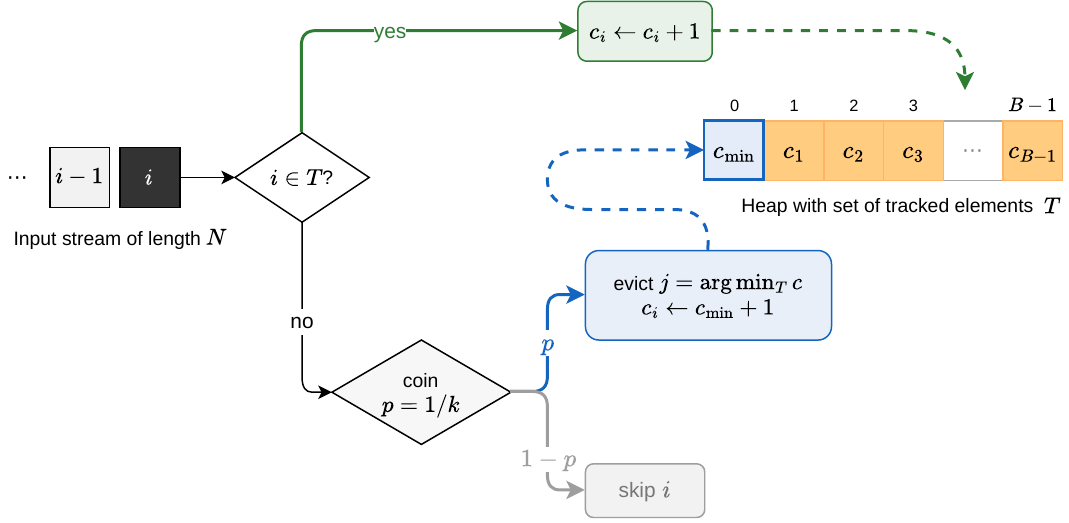}
    \caption{Visualization of the update mechanism of SpaceR.}
    \label{fig:spaceR}
\vspace{-0.2cm}
\end{figure}

\textbf{The algorithm.} SpaceR can be thought of as a randomized generalization of SpaceSaving. Similar to the \emph{Sample and Hold} approach by~\citet{SampleAndHold}, it relies on the idea of sampling but differs in its memory allocation to avoid an unbounded and data-dependent memory footprint, as exhibited by \emph{Sample and Hold}.
To do so, SpaceR maintains a similar heap of elements together with associated counts. When it processes a new element, it first checks if the element is already in the heap. If so, it increments its associated counter. Otherwise, it randomly decides whether to process the element. Namely, with probability $p$, it inserts the element using the SpaceSaving replacement strategy: it evicts the element in the heap with the minimum counter $c_{\min}$, and initializes the new element with count $c_{\min} + 1$. With probability $1-p$, it skips the element (see ~\cref{alg:spacer_insert} and ~\cref{fig:spaceR} for pseudocode and visualization). On a query, if the element is stored in the heap, it returns its associated counter. Otherwise, it returns $0$.

\begin{algorithm}[]
  \caption{SpaceR Insert Algorithm}
  \label{alg:spacer_insert}
  \begin{algorithmic}[1]
    \REQUIRE Stream of $n$ distinct items, number of heavy hitters $k$, space budget $B = \Theta(k)$, probability parameter $p = \frac{1}{k}$
    \STATE $H \gets$ heap of size $B$
    \STATE $T \gets \{\}$ set of elements tracked in heap
    \STATE $c_x \gets 0$ for all counts in heap
    \FOR{each stream element $i$}
      \IF{$i \in T$}
        \STATE $c_i \gets c_i + 1$
      \ELSE
        \STATE With probability $p$:
            \STATE \hspace{1em} Let $j$ be the element with minimum count in $H$
            \STATE \hspace{1em} $c_i \gets c_j + 1$
            \STATE \hspace{1em} Remove $j$ from $T$ and $H$
            \STATE \hspace{1em} Add $i$ to $T$ and $H$ with count $c_i$
      \ENDIF
    \ENDFOR
  \end{algorithmic}
\end{algorithm}
\textbf{The probability $p$.}  We propose to set the probability $p$ inversely proportional to the number of heavy hitters we would want to detect, i.e., $p=\frac{1}{k}$, as this is directly linked to their frequency in a Zipfian stream. Namely, the effect of the probability $p$ is that it makes SpaceR run almost as if we were running SpaceSaving on a sub-sampled stream, in which rare elements appear less often while preserving most appearances of heavy hitters.~\footnote{We note that one could consider processing every element of the stream with probability $p$, not just the ones that would trigger the SpaceSaving replacement strategy. Based on existing theoretical analyses, it is unclear whether this would lead to similar performance gains as SpaceR.} 

The intuition for our choice of $p$ is that heavy hitters are more likely to end up in the heap as they appear more often. More broadly, this choice of $p$ filters out the high-volume noise of rare elements in the long tail while maintaining enough sensitivity to capture all elements that genuinely belong to the top-$(B/2)$. Once tracked, they benefit from two mechanisms: first, all subsequent appearances are counted while they are in the heap (no coin toss for tracked elements); second, new insertions occur only with probability $p$, allowing tracked elements to persist longer in the heap. This gives heavy hitters more time to accumulate counts and distance themselves from $c_{\text{min}}$. Moreover, as the gap between a heavy hitter's counter $c_i$ and $c_{\text{min}}$ grows, it becomes increasingly difficult to evict that element from the heap. 

Unlike SpaceSaving, which tracks every occurrence, SpaceR discards many rare element appearances before they enter the data structure. As we demonstrate in ~\cref{experiments}, this approach performs remarkably well in practice. SpaceR not only outperforms SpaceSaving at retrieving heavy hitters, but also approximates their frequencies with near-perfect accuracy even for space budgets as small as $4k$, where $k$ is the number of heavy hitters of interest. Beyond heavy-hitter detection, this level of accuracy makes SpaceR a practical tool for frequency estimation. 

%% file: Chapters/05_Experiments.tex
\section{Experiments} \label{experiments}
\begin{figure}
    \centering
    \begin{subfigure}{\textwidth}
        \centering
        \includegraphics[width=\textwidth]{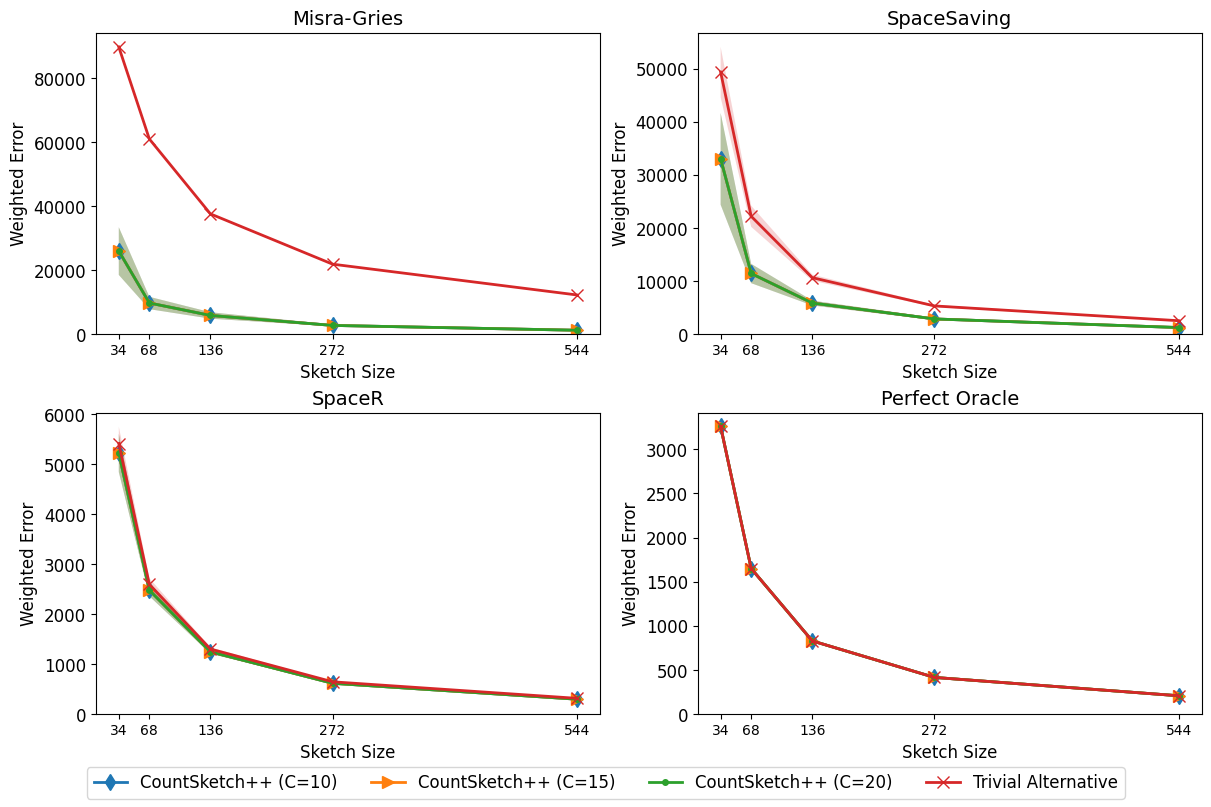}
        \caption{Synthetic data}
        \label{fig:exp_1_synthetic}
    \end{subfigure}
    \vspace{1em} 
    \begin{subfigure}{\textwidth}
        \centering
        \includegraphics[width=\textwidth]{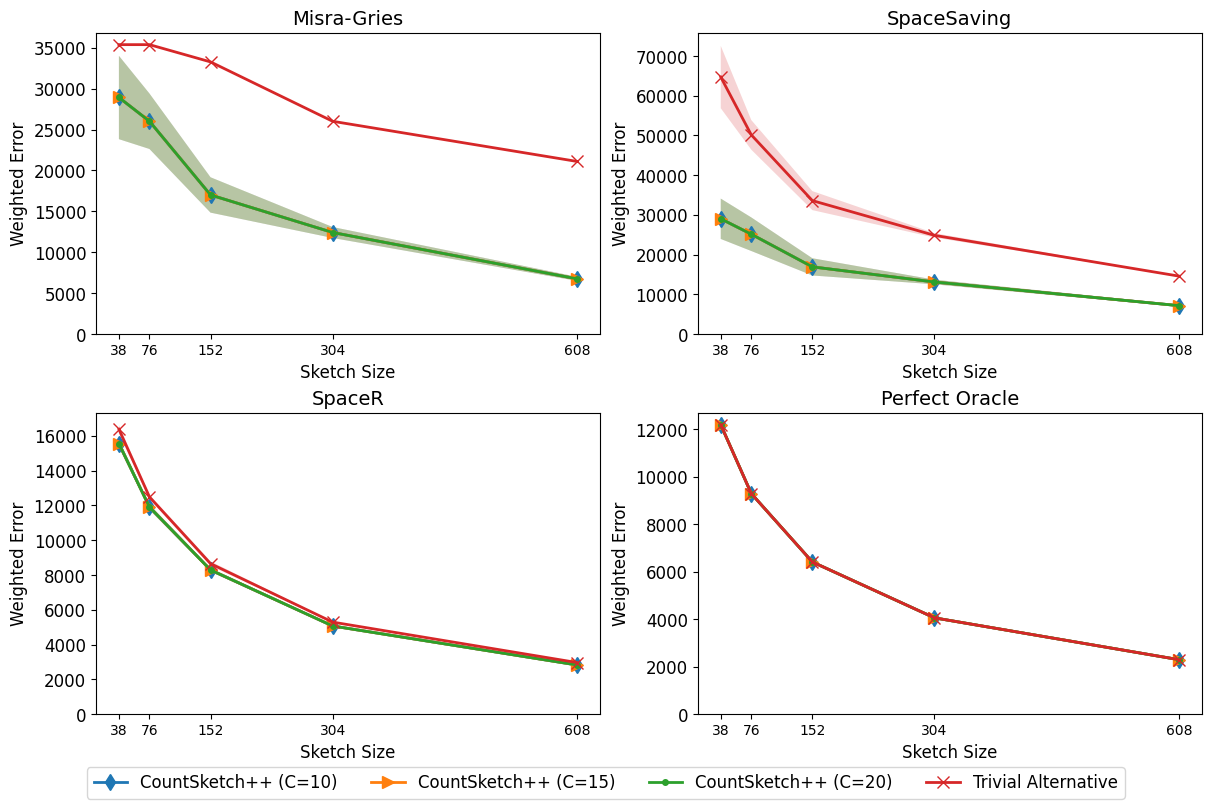}
        \caption{CAIDA dataset}
        \label{fig:exp_1_caida}
    \end{subfigure}
    \caption{Comparison of weighted errors for different oracles in the first pass, combined with either the Trivial Alternative or a CS++ in the second. Results are listed by oracle. In the second pass, the learned CS++ dedicates $B/2$ space to track exact counts for the heavy hitters and employs a CS++ sketch with the remaining $B/2$ space. Note that the Trivial Alternative only ever stores the heavy hitters. That is, it only uses a fraction of the available space budget in the second pass.}
    \label{fig:exp_1}
\vspace{-0.2cm}
\end{figure}

Our experiments are structured into two parts. First, we investigate the effectiveness of the Trivial Alternative relative to learned CS++ when using different heavy-hitter oracles (\cref{fig:exp_1}). Second, we evaluate the performance of our one-pass SpaceR algorithm against established two-pass, learning-augmented baselines to see if a single pass can provide comparable accuracy (\cref{fig:exp_2}).

We ran all experiments five times with different random seeds and report the average results and standard deviation. In addition, we varied the threshold parameter $\tau$, which decides whether CS++ outputs zero or instead returns the estimate from the large sketch table, since the theoretical analysis defines asymptotic values for $\tau$. Specifically, we tested three different values for the factor $C$ in the asymptotic bound of $\tau$. Unless the results differ significantly based on the choice of $C$, we will only report the results for one of the constants tested. 
We provide the results for all constants in \cref{appendix:experiments}. Following previous work~\cite{MIT}, space complexity is measured in words of memory. 

\textbf{Datasets.} We compare our results to prior work on synthetic Zipfian data and the CAIDA dataset \cite{CAIDA_DATA}. CAIDA is a real-world dataset collection of anonymized internet traces. We ran our experiments on the first minute of the recordings from January 2019, which contains 29,922,875 Tier-1 ISP packets ($N$) from 326,358 different source IP addresses ($n$). For the synthetic data, we generate a Zipfian data stream with $n= 10^6$ such that the $i^{th}$ element has a frequency of $f_i = \left\lceil \frac{n}{i} \right\rceil$.

\textbf{Space allocation.}  
We evaluate all methods under a fixed total space budget $B$ to ensure that all sketches allocate the same amount of memory. For algorithms that rely on predictions, this includes the oracle as well, which is different from many of the prior works that do not account for the space to build and store the oracle. 
Since we implement the oracle using key-value pairs, each entry requires two units of memory: one to store the identity (key) of the heavy hitter and another for its frequency count. Consequently, the full budget $B$ accommodates up to $B/2$ counters. Following the allocation strategy in~\citet{MIT, ERRORDEF}, the learned CS++ divides this budget equally during the estimation phase: $B/2$ is assigned to track heavy hitters, while the remaining $B/2$ is dedicated to the sketching component. In contrast, the Trivial Alternative stores only the heavy hitters and omits the sketch entirely. Consequently, the Trivial Alternative utilizes only a fraction of the available budget during the second pass, making it strictly more space-efficient than the learned CS++.

\subsection{Trivial Alternative vs. learned CS++}\label{sec:experiment1}
The first experiment evaluates the interplay between the choice of oracle in the first part and the estimation mechanism in the second part. 
We employ the two-pass learned framework from~\citet{ERRORDEF}, where the first pass is dedicated to constructing a heavy-hitter oracle and the second pass performs frequency estimation with either a learned CS++ or the Trivial Alternative. 
We do not compare against (learning-augmented versions of) CountMin and CountSketch in the second pass, as they perform worse in theory \cite{MIT} and practice \cite{MIT, ERRORDEF}.

\textbf{Results.} \Cref{fig:exp_1} illustrates the performance of the evaluated approaches across both datasets. When paired with high-quality predictions from a perfect oracle or SpaceR, the Trivial Alternative achieves weighted errors nearly identical to the learned CS++ across both datasets. This parity occurs due to the thresholding logic of CS++: when an oracle correctly identifies the primary heavy hitters, the remaining elements have actual frequencies so low that CS++ defaults to outputting zero. This makes the additional CS++ sketch in the learned CS++ unnecessary. 

Conversely, a performance gap emerges when using lower-quality oracles. In the case of Misra-Gries, which exhibits significant error in identifying heavy hitters, the learned CS++ outperforms the Trivial Alternative. This divergence occurs because the sketching component in CS++ is able to partially mitigate the oracle's mispredictions. For SpaceSaving, the Trivial Alternative remains competitive with learned CS++ on the synthetic data but performs worse on the CAIDA dataset.

\subsection{One-pass SpaceR vs. two-pass learned frameworks} \label{exp_2}
Next, we evaluate whether our proposed one-pass algorithm, SpaceR, can match the performance of SOTA learning-augmented methods that require two passes. Specifically, we compare the direct one-pass versions of SpaceR and SpaceSaving, which return estimates straight from their internal counters, against the two-pass configurations established in Experiment 1 (see \cref{sec:experiment1}) (pairing the oracles with either the Trivial Alternative or learned CS++).

\begin{figure}
  \begin{centering}
    \centerline{\includegraphics[width=0.95\textwidth]{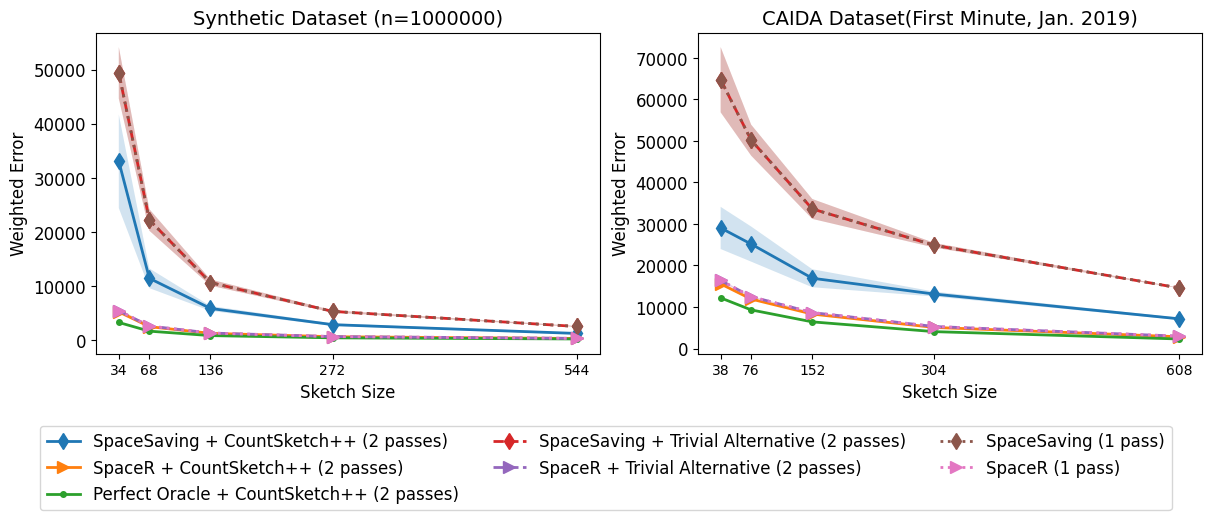}}
    \caption{
      Comparison of weighted error between the best-performing approaches over one and two passes for the synthetic and the CAIDA datasets. The results are shown for the threshold in CS++ with $C=15$.
    }\label{fig:exp_2}
    \centerline{\includegraphics[width=0.95\textwidth]{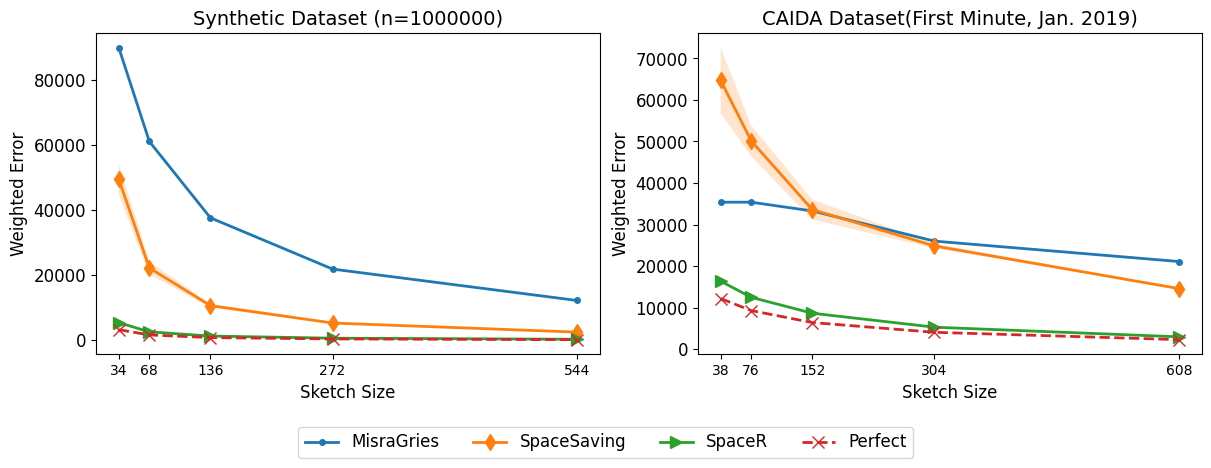}}
    \caption{
      Comparison of different oracles on how well they perform for frequency estimation. Estimates are retrieved by outputting the counts associated with the top-$k$ elements for the heavy hitters and outputting zero for all remaining elements.  
    }\label{fig:oracle_comparison}
  \end{centering}
\end{figure}

\textbf{Results.} As shown in~\cref{fig:exp_2}, all configurations utilizing SpaceR achieve performance parity even with the most powerful two-pass baseline, that is, the perfect oracle combined with learned CS++. Notably, this result holds true even for the direct one-pass version of SpaceR. This demonstrates that SpaceR not only identifies the heavy hitters but also recovers their frequencies with near-perfect accuracy in a single pass, even under highly constrained space budgets.

Further analysis in \cref{fig:oracle_comparison} reveals that SpaceR achieves stable performance at a space budget as low as $B=4\log(n)$ in the experiments. It manages to recover reasonable frequency estimates for half of the elements tracked. 

Across both datasets, SpaceR consistently outperforms other oracles tested. The comparison with the SpaceSaving algorithm is especially interesting as SpaceR is a randomized version of the former: The experiments indicate that the ``coin toss'' mechanism significantly improves both heavy-hitter identification and frequency recovery compared to the deterministic SpaceSaving algorithm. 

Specifically, heavy hitters profit from two mechanisms in SpaceR: As the algorithm filters insertions of untracked elements with probability $p$, it evicts counters much less frequently and therefore stabilizes the heap. This allows tracked elements to persist in the heap longer, providing them the necessary time to accumulate counts and widen the gap between their counter $c_i$ and the minimum counter $c_{\text{min}}$. As this gap grows, it becomes increasingly difficult to evict these elements, allowing SpaceR to ``protect'' the counts of heavy hitters and achieve the near-perfect recovery observed in the experiments.
\renewcommand{\topfraction}{0.9}
\renewcommand{\bottomfraction}{0.9}
\renewcommand{\textfraction}{0.2}

\FloatBarrier

%% file: Chapters/06_Conclusion.tex
\section{Conclusion} \label{conclusion}
In this work, we have re-examined the design space of frequency estimation in skewed data streams and provided a new perspective on the trade-off between algorithmic complexity and predictive utility. We demonstrated that unlearned algorithms can match the performance of their learned counterparts with only a logarithmic increase in space – $\Theta(B \log B)$ vs $\Theta(B)$ – and showed that the more complex structure of current learned frameworks can be replaced by simpler approaches on Zipfian data.

Our proposed algorithm, SpaceR, empirically matches the performance of SOTA two-pass, learning-augmented algorithms even under tightly constrained space budgets, without requiring external predictions or multiple passes over the data. We validated this on both synthetic Zipfian data and the real-world CAIDA dataset.

While our results are promising, they suggest several avenues for future inquiry. First, a formal theoretical characterization of the error bounds for randomized admission policies like SpaceR remains an open challenge. Second, while our study focused on the prevalent Zipfian distribution, it is unclear how these results generalize to distributions with lower skewness. 

Ultimately, our results show that the cost of generating predictions can exceed the benefit they provide and suggest that algorithmic innovations may better exploit problem structure than learned predictions. By demonstrating that one-pass, randomized algorithms can match or exceed the performance of complex learned systems, we provide a pathway for resource-efficient data processing in constrained environments.

\FloatBarrier

%% file: Chapters/appendix_improved_bounds.tex
\counterwithin{theorem}{section}
\section{Improved analysis of CountSketch++ without predictions} \label{appendix:improved_analysis}
In this section, we improve the analysis for the upper bound of the weighted error for the CountSketch++ algorithm without predictions.
The intuition follows the proof technique in~\citet{MIT}. 
More specifically, we prove the following theorem as stated in \cref{theory1}:

\theoImprov*

\begin{figure}[ht]
 \centering
\begin{tikzpicture}[scale=1.5]
\draw[->] (0,0) -- (7,0) node[right] {$i$};
\draw[->] (0,0) -- (0,5) node[above] {};
\draw (0.8,0.1) -- (0.8,-0.1) node[below] {$\frac{A}{2\tau}$};
\draw (3.2,0.1) -- (3.2,-0.1) node[below] {$\frac{2A}{\tau}$};
\draw (1.6,0.1) -- (1.6,-0.1) node[below] {$\frac{A}{\tau}$};
\draw (-0.1,0.65) -- (0.1,0.65);
\node at (-0.25,0.65) {$\tau$};
\draw[dashed] (0.8,0) -- (0.8,5);
\draw[dashed] (3.2,0) -- (3.2,5);
\draw[dashed] (1.6,0) -- (1.6,0.65);
\draw[dashed] (7,0.5) -- (0,0.65);
\draw[thick] plot[domain=0.2:7, samples=100] (\x, {1/\x});
\draw (0.5,3.5) circle (0.2) node {1};
\draw (2.,3.5) circle (0.2) node {3};
\draw (5.,3.5) circle (0.2) node {2};

\node[draw, rectangle, fill=TUMblue!15, align=center, text width=4cm] at (1.8,4.6) 
    {\footnotesize  \textbf{Heavy Hitters}\\ CountSketch++ outputs estimate from big table w.h.p.};
\draw[->, thick] (1.8,4.) -- (0.6,3.7) ;

\node[draw, rectangle, fill=TUMblue!15, align=center, text width=3cm] at (5,2.) 
    {\footnotesize  \textbf{Rare Elements}\\ CountSketch++ outputs zero w.h.p.};

\node[draw, rectangle, fill=TUMblue!15, align=center, text width=3cm] at (2,2) 
    {\footnotesize  \textbf{Medium Frequency}\\ Assume worst case, i.e., CountSketch++ outputs zero};

\node[rotate=90] at (-0.5,2.5) {$f_i$};
\end{tikzpicture}
\caption[Element Partitions for CountSketch++]{Intuitively, the CountSketch++ algorithm splits the elements into three groups depending on their frequency relative to the threshold $\tau$ used to filter rare elements.}
\label{fig:element_groups}
\end{figure}
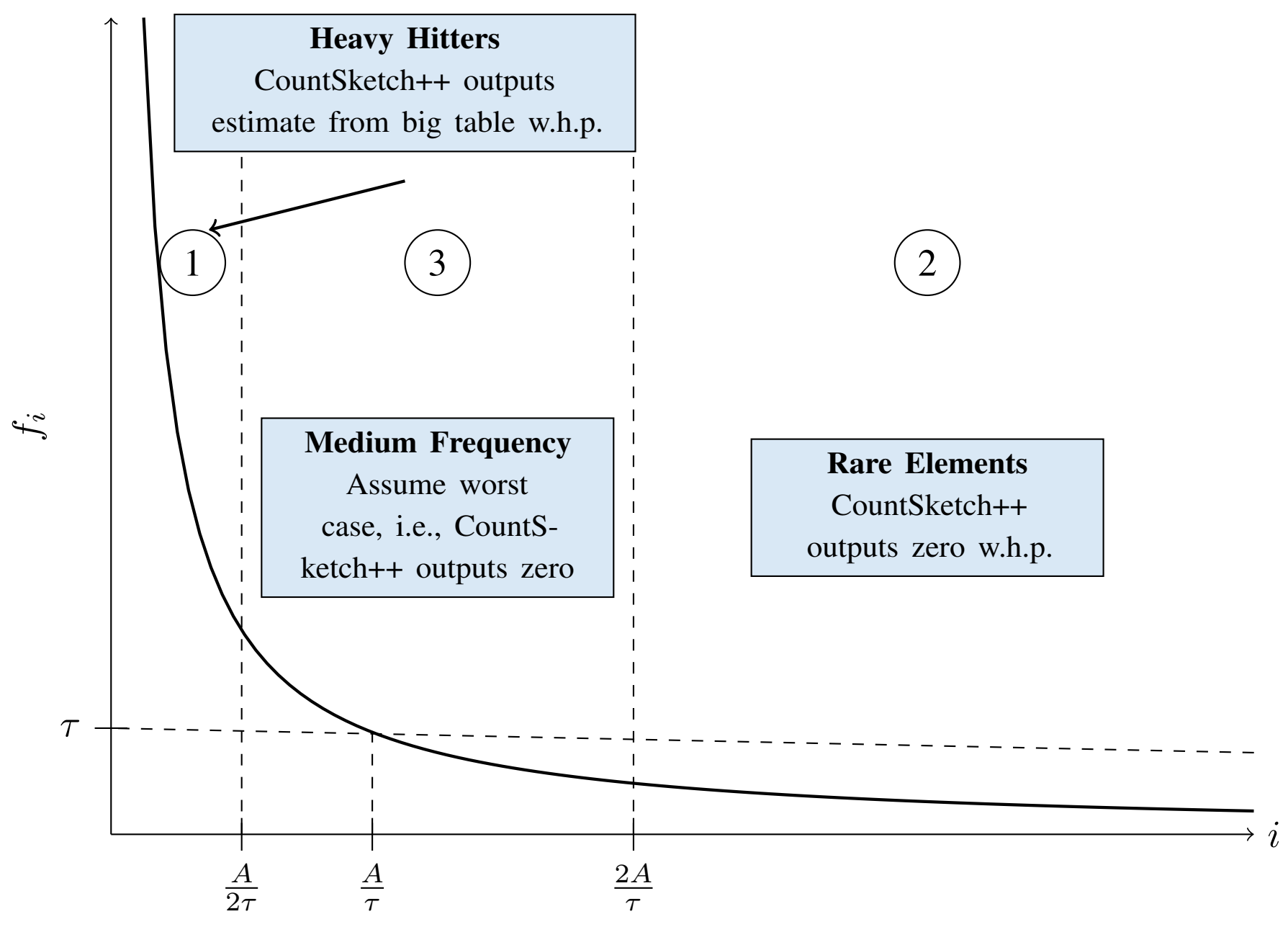

%
\textbf{Proof intuition.} 
Intuitively, the algorithm partitions the elements into three groups: (cf.~\cref{fig:element_groups}): Group one contains the heaviest elements. Group two comprises the rarest elements.  
Finally, group three holds all elements in between whose frequencies are close to the threshold $\tau$ used to determine whether an element is heavy or not during the filtering step.
Reducing the number of elements in group three,
we can reduce the expression from $ \bigO(\frac{N}{\log^2(n)}\cdot \frac{\log B + \mathrm{poly}(\log \log n) }{B})$  to $\bigO(\frac{N}{\log^2(n)}\cdot \frac{\log B }{B})$.
To shrink group three, we keep the threshold $\tau$ as suggested in \cite{MIT} but tighten the boundaries from originally $\left(\frac{A}{\tau}, \frac{B}{(\log \log n)^4}\right)$ to $\left(\frac{A}{2\tau}, \frac{2A}{\tau}= \frac{2 B}{\bigO(\log \log n)}\right)$.
This improves the original error bound by a factor of $\frac{1}{\mathrm{poly}(\log \log n)}$, where $n$ is the number of distinct elements in the stream.  \\

Given a threshold of $\tau= \frac{A \cdot \bigO(\log \log n)}{B}$, we show that for any element $i: f_i \leq \frac{\tau}{2}$, the algorithm outputs zero with very high probability. 
For any element $i: f_i \geq 2\tau$, the algorithm outputs the estimate in the big sketch table w.h.p. 
For any element in between these two thresholds, we assume the worst case, that is, that the algorithm outputs zero. 
Remember that outputting zero incurs an error of $\frac{A}{i}$, whereas outputting the estimate from the big sketch table results in an error of $\frac{A}{B}$ (cf. \cite{MIT}, Theorem C.4).  \\ 

\begin{proof}
We are going to use the following theorem as stated and proved in~\citet{MIT}: 

\begin{theorem}\label{thm_MIT_5}
Let $\hat{f}_i$ be the estimate of the $i^{th}$ frequency given by a 3 x $\frac{B'}{3}$ CountSketch table, where $B' \geq \log(n)$. There exists a universal constant C such that the following inequality holds: 
\[\forall t \geq 3/B', \qquad Pr[|f_i-\hat{f}_i| \geq t] \leq C' \left( \frac{\log(tB')}{tB'}\right)^2\]
\end{theorem}

First, we decompose the total expected weighted error, {$\mathbf{E_{total}}$}, into three parts. Each part represents the error incurred by the respective group of elements, 
where group one contains Heavy Hitters, group two represents rare elements, and group three the elements in between (cf.~\cref{fig:element_groups}): 
\[E_{total} = \frac{1}{N} \sum_{i \in [n]} f_i \cdot |\hat{f_i}- f_i| = \frac{1}{N} (E_1 + E_2 + E_3) \]

For all groups of elements, let us decompose the expected weighted error $\mathbf{E_x, x \in \{1,2,3\}}$ they contribute to the final error $E_{total}$ into two parts $E_S$ and $E_F$. 
$\mathbf{E_S}$ describes the error obtained if we output the estimate in the sketch. $\mathbf{E_F}$ is the error incurred by outputting zero.
Weighting each part, $E_S$ and $E_F$, by the probability that these events occur (i.e., outputting zero vs. outputting the estimate from the big sketch table) leads to: 
\[E_x =  \sum_{i \in [G_x]}\left( p_S(G_x) \cdot  E_S + p_F(G_x) \cdot  E_F\right)\]
where $\mathbf{G_x, x \in \{1,2,3\}}$ is the group of elements under consideration, $\mathbf{p_S(G_x)}$ is the probability that the algorithm outputs the estimate from the big table for elements in group $G_x$, and $\mathbf{p_F(G_x)}$ is the probability that the algorithm outputs zero. 
Additionally, we know that for all groups: 
\[E_S = \E[f_i \cdot|f_i- \tilde{f}_i|] = \frac{A}{i} \cdot \frac{A}{B} = \frac{A^2}{iB}\]
\[E_F = \E[f_i \cdot|f_i- 0|] = \frac{1}{i} \cdot \frac{A}{i} = \frac{A}{i^2}\]
Since for all $i\leq B$ it holds that $E_S = \frac{A}{i} \cdot \frac{A}{B} \leq  \frac{A^2}{i^2} = E_F$, it is desirable to output the estimate from the big CountSketch table $\tilde{f}_i$ for these elements. 
For all other elements, it follows that it is preferable to output zero.  \\ 

\textbf{Case 1:} This case deals with all elements $i$, such that 
\[i\leq\frac{A}{2\tau} = \frac{B}{2 \log \log n} \xrightarrow{} f_i= \frac{A}{i}\geq 2\tau = \frac{2 A \log \log n}{B}\]
For these elements, it is desirable that the algorithm outputs the estimate from the big sketch table. 
Accordingly, the bad scenario is that the algorithm labels these frequent elements as rare and outputs zero, since 
\[\forall i \leq \frac{A}{2 \tau}: \qquad f_i = \frac{A}{i} \geq \frac{2 A\log \log n}{B} > \frac{A}{B}\] 
, where $\frac{A}{B}$ is the expected additive error incurred by a CountSketch table with $B$ columns. 
To upper bound the error $E_1$, we aim to minimize the upper bound of the bad scenario, i.e., the probability that the algorithm outputs zero, $p_F$. 
Let $\hat{f}_i^{(k)}$ be the frequency estimate of the $i^{th}$ element given by the $k^{th}$ small CountSketch table.
In the following, $\lTilde$ denotes inequality up to a constant factor. 
The probability that a single estimate from a small table $\hat{f}_i^{(k)}$ is smaller than $\tau$ is:
\begin{flalign*}
\Pr[\hat{f}_i^{(k)} \leq \tau ] &= \Pr\left[-\hat{f}_i^{(k)} \geq -\tau\right] &\quad& \textit{Multiply both sides by -1}\\
&= \Pr\left[f_i - \hat{f}_i^{(k)} \geq f_i - \tau\right] &\quad& \textit{Add $f_i$ to both sides} \\
&\leq \Pr\left[|f_i - \hat{f}_i^{(k)}| \geq f_i - \tau\right] &\quad& \textit{Take the absolute value of the left side}  \\
&\leq \Pr\left[|f_i - \hat{f}_i^{(k)}| \geq 2 \tau - \tau\right] &\quad& \textit{Lower bound $f_i$ by $2\tau$, the minimum frequency in group 1} \\
&\leq \Pr\left[|f_i - \hat{f}_i^{(k)}| \geq \tau\right] \\
\end{flalign*}
Given that term, we can now apply~\cref{thm_MIT_5}:
\begin{flalign*}
\Pr\left[|f_i - \hat{f}_i^{(k)}| \geq \tau\right]
&\leq C'\left(\frac{\log(\tau\mathcal{B'})}{\tau\mathcal{B'}}\right)^2 &\quad& \textit{Apply~\cref{thm_MIT_5}} \\
&\leq C'\left(\frac{\log(\frac{C  A \cdot \log\log n}{B} \cdot \frac{B}{2T})}{\frac{C A \cdot \log\log n}{B} \cdot \frac{B}{2T}}\right)^2 &\quad& \textit{Plug in values: $\tau= \frac{A \cdot \bigO(\log \log n)}{B}$ and $B'= \frac{B}{2T}$} \\
&\leq C'\left(\frac{\log(\frac{C  A \cdot\log\log n}{B} \cdot \frac{B}{2C''\log\log n})}{\frac{C A \cdot\log\log n}{B} \cdot \frac{B}{2C''\log\log n}}\right)^2 &\quad& \textit{Plug in value for $T, T= \bigO(\log \log n)$} \\
&\leq C'\left(\frac{2C'' \log\left(\frac{AC}{2C''}\right)}{AC}\right)^2 &\quad& \textit{Simplify terms where possible} \\
&\leq 4C'C''^2\left(\frac{\log\left(\frac{AC}{2C''}\right)}{AC}\right)^2 &\quad& \textit{Factor out 2C''} \\
&\lTilde \left(\frac{\log(AC)}{AC}\right)^2 \lTilde \frac{1}{8}&\quad& \textit{Omit constant parts and upper bound by $\frac{1}{8}$} \\
\end{flalign*}
We have now shown, that the probability that a single small CountSketch table in the algorithm labels a heavy element as rare is at most $\frac{1}{8}$.
Next, we want to upper bound the probability that the filtering step as a whole labels a heavy element as rare. 
This happens if at least half of the $T$ small tables return an estimate smaller than $\tau$.
Let $m$ be the number of small tables with estimates smaller than $\tau$. For the CountSketch++ to label an element as rare, it must hold that  $m \geq \frac{T}{2}$.
There are $\binom{T}{m}$ different combinations to choose $m$ from the $T$ estimates.
Given that the estimates are independent, and the probability that a single estimate from a small table is smaller than $\tau$ is upper bounded by $\frac{1}{8}$ as stated above, we get: 
\begin{flalign*}
    p_F(G_1) &= \sum_{m=T/2}^T \binom{T}{m} \cdot Pr[\hat{f}_i^{(k)} \leq \tau ]^m \cdot Pr[\hat{f}_i^{(k)} > \tau ]^{T-m} \\
    &\leq \sum_{m=T/2}^T \binom{T}{m} \cdot Pr[\hat{f}_i^{(k)} \leq \tau ]^m &\quad& \textit{Omit $Pr[\hat{f}_i^{(k)} > \tau ]^{T-m}$} \\ 
    & \leq \frac{T}{2} \cdot \binom{T}{T/2} \cdot Pr[\hat{f}_i^{(k)} \leq \tau ]^{T/2} &\quad& \textit{Upper bound the single addends $\left(\textit{max. for } m= \frac{T}{2}\right)$} \\ 
    & \leq \frac{T}{2} \cdot \left( \frac{2eT}{T} \right)^{T/2} \cdot Pr[\hat{f}_i^{(k)} \leq \tau ]^{T/2} &\quad& \textit{Upper bound binom. coefficient by $\binom{n}{k} \leq \left(\frac{e n}{k}\right)^k$}\\
    & \leq \frac{T}{2} \cdot \left( 2e \right)^{T/2} \cdot Pr[\hat{f}_i^{(k)} \leq \tau ]^{T/2} &\quad& \textit{Simplify terms}\\
    &\lTilde \frac{T}{2} \cdot \left( 2e \right)^{T/2} \cdot \left(\frac{1}{8}\right)^{\frac{T}{2}} &\quad& \textit{Plug in upper bound for $Pr[\hat{f}_i^{(k)} \leq \tau ]$}\\
    &\lTilde \frac{T}{2} \cdot \left( 2e \right)^{T/2} \cdot e^{-\ln(8)\frac{T}{2}} & \quad & \textit{Use trick that $\frac{1}{8}= e^{-\ln(8)} $}\\
    &\lTilde \frac{T}{2} \cdot \left( 2e \right)^{T/2} \cdot e^{-\Theta(T)} &\quad& \textit{Rewrite $e^{-\ln(8)\frac{T}{2}}$ as $e^{-\Theta(T)}$}\\
    &\lTilde e^{-\Theta(T)}  &\quad& \textit{Omit further constant parts of term} \\
    &\lTilde e^{-\Theta(\log \log n)}  &\quad& \textit{Plug in value for $T$} \\
    &\lTilde \frac{1}{\log^C(n)} &\quad &\textit{Rewrite term; plug in constant $C$ for $\Theta$} \\
    & \lTilde \frac{1}{\log^{100}(n)} &\quad &\textit{Plug in concrete value for C}
\end{flalign*}

Plugging in the obtained upper bound for $p_F(G_1)$, the probability that the bad scenario occurs, i.e., the algorithm outputs zero, and upper-bounding the probability for the good scenario $p_S(G_1)$ by one, we get: 
\begin{flalign*}
E_1 &= \sum_{i=1}^{\frac{A}{2\tau}} \E[f_i \cdot|f_i- \hat{f}_i|] \\
&= \sum_{i=1}^{\frac{A}{2\tau}} \left( p_F(G_1) \cdot  E_F +p_S(G_1) \cdot  E_S\right) \\
&\lTilde \sum_{i=1}^{\frac{A}{2\tau}} \left(p_F(G_1) \cdot \frac{A^2}{i^2} + p_S(G_1) \cdot \frac{A^2}{iB}  \right)  &\quad & \textit{Plug in $E_S = \frac{A^2}{iB}$ and $E_F = \frac{A^2}{i^2}$} \\
&\lTilde \sum_{i=1}^{\frac{A}{2\tau}} \left( \frac{1}{\log^{100}(n)} \cdot \frac{A^2}{i^2} + \frac{A^2}{iB} \right)  &\quad & \textit{Plug in obtained upper bound for $p_F(G_1)$ } \\
&& \quad & \textit{and upper bound $p_S(G_1)$ by one} \\ 
&\lTilde \frac{A^2}{\log^{100}(n)} \sum_{i=1}^{\frac{A}{2\tau}} \frac{1}{i^2} + \frac{A^2}{B} \sum_{i=1}^{\frac{A}{2\tau}} \frac{1}{i}  &\quad & \textit{Factor out constants}\\
&\lTilde \frac{A^2}{\log^{100}(n)} \int_1^{\frac{A}{2\tau}} \frac{1}{x^2} \, dx + \frac{A^2}{B}  \cdot \log\left(\frac{A}{2\tau} \right) &\quad & \textit{Upper bound left sum by integral and right sum}\\
& & \quad & \textit{by harmonic series $H(n) = \sum_{x=1}^n \frac{1}{x}\lTilde \log(n)$} \\
&\lTilde \frac{A^2}{\log^{100}(n)} \cdot \left(1- {\frac{2\tau}{A}}\right) + \frac{A^2}{B} \cdot \log\left(\frac{A}{2\tau} \right)&\quad & \textit{Calculate integral}\\
&\lTilde \frac{A^2}{\log^{100}(n)} \cdot \left( 1 - {\frac{2 \log \log n}{B}}\right) + \frac{A^2}{B} \cdot \log\left(\frac{B}{2\log \log n} \right) &\quad & \textit{Plug in value for $\tau$}\\
&\lTilde A^2 \left(\frac{B-2 \log \log n}{\log^{100}(n) B} + \frac{\log B}{B} - \frac{\log (2 \log \log n)}{B}\right) &\quad & \textit{Factor out$A^2$ and pull apart logarithmic term}\\
&=\bigO\left(\frac{A^2 \log B}{B}\right) &\quad & \textit{Only keep dominating part}
\end{flalign*}

\textbf{Case 2:} This case deals with all elements $i$, such that
\[i\geq \frac{2A}{\tau} = \frac{2B}{\log \log n} \xrightarrow{} f_i = \frac{A}{i}\leq \frac{\tau}{2} = \frac{A\log \log n}{2B}\]
Note that not for all elements in this group, the best and worst cases are the same: For some elements $i$ such that $\frac{2B}{\log \log n} \leq i \leq B$, it is indeed still preferable to output the estimate from the big sketch table.
However, for most elements in group two, the best and worst case scenarios are just inverse to the first case:
It is desirable that the algorithm outputs zero, while the bad scenario is that the algorithm outputs the estimate from the big sketch table. \\ 

Following the same reasoning as in the first case, we can upper bound the error $E_2$ by upper bounding the bad event, i.e., the probability that the algorithm outputs the estimate from the big sketch table, $p_S$.
To do so, let us first consider the probability that a single estimate from the small tables is larger than $\tau$:
\begin{flalign*}
\Pr[\hat{f}_i^{(k)} \geq \tau ] &= \Pr\left[\hat{f}_i^{(k)}-f_i \geq  \tau - f_i\right]  &\quad& \textit{Subtract $f_i$ from both sides} \\
&\leq \Pr\left[|\hat{f}_i^{(k)}-f_i| \geq  \tau -f_i \right]  &\quad&  \textit{Take the absolute value of the left side}  \\
&\leq \Pr\left[|\hat{f}_i^{(k)}-f_i | \geq \tau - \frac{\tau}{2}\right] &\quad& \textit{Upper bound $f_i$ by $\frac{\tau}{2}$, the max. frequency in group 2} \\ 
&\leq \Pr\left[| \hat{f}_i^{(k)}-f_i | \geq \frac{\tau}{2}\right] \\
\end{flalign*}
Like in the first case, we can now apply~\cref{thm_MIT_5}:
\begin{flalign*}
\Pr\left[|f_i - \hat{f}_i^{(k)}| \geq \frac{\tau}{2}\right] &\leq C'\left(\frac{\log(\frac{\tau}{2}B')}{\frac{\tau}{2}B'}\right)^2 &\quad& \textit{Apply~\cref{thm_MIT_5}} \\
&\leq C'\left(\frac{\log(\frac{CA\log\log n}{2B} \cdot \frac{B}{2T})}{ \frac{CA \cdot \log\log n}{2B} \cdot \frac{B}{2T}}\right)^2 &\quad& \textit{Plug in values: $\tau= \frac{A\bigO(\log \log n)}{B}$ and $B'= \frac{B}{2T}$} \\
&\leq C'\left(\frac{\log(\frac{CA\log\log n}{2B} \cdot \frac{B}{2C''\log\log n})}{ \frac{CA \cdot \log\log n}{2B} \cdot \frac{B}{2C''\log\log n}}\right)^2 &\quad& \textit{Plug in value for $T, T= \bigO(\log \log n)$} \\
&\leq C'\left(\frac{4C'' \log\left(\frac{AC}{4C''}\right)}{AC}\right)^2  &\quad& \textit{Simplify terms where possible} \\
&\leq 16C'C''^2\left(\frac{\log\left(\frac{AC}{4C''}\right)}{AC}\right)^2 &\quad& \textit{Factor out constants} \\
&\lTilde \left(\frac{\log(AC)}{AC}\right)^2 \lTilde \frac{1}{8} &\quad& \textit{Omit irrelevant constants and upper bound by $\frac{1}{8}$} \\
\end{flalign*}
Now that we have upper-bounded the probability that a single small table fails and labels a rare element as heavy, we can determine the probability that the filtering step as a whole labels a rare element as heavy.
That happens if at least half of the $T$ small tables return an estimate larger than $\tau$. Let $m$ be the number of small tables with estimates larger than $\tau$.
The steps follow the same logic as in case one:
\begin{flalign*}
    p_S(G_2) &= \sum_{m=T/2}^T \binom{T}{m} \cdot Pr[\hat{f}_i^{(k)} \geq \tau ]^m \cdot Pr[\hat{f}_i^{(k)} < \tau ]^{T-m} \\
    &\leq \sum_{m=T/2}^T \binom{T}{m} \cdot Pr[\hat{f}_i^{(k)} \geq \tau ]^m &\quad& \textit{Omit $Pr[\hat{f}_i^{(k)} < \tau ]^{T-m}$} \\ 
    & \leq \frac{T}{2} \cdot \binom{T}{T/2} \cdot Pr[\hat{f}_i^{(k)} \geq \tau ]^{T/2} &\quad& \textit{Upper bound the single addends $\left(\textit{max. for } m= \frac{T}{2}\right)$} \\ 
    & \leq \frac{T}{2} \cdot \left( \frac{2eT}{T} \right)^{T/2} \cdot Pr[\hat{f}_i^{(k)} \geq \tau ]^{T/2}  &\quad& \textit{Upper bound binom. coefficient by $\binom{n}{k} \leq \left(\frac{e n}{k}\right)^k$}\\ 
    &\leq \frac{T}{2} \cdot \left( 2e \right)^{T/2} \cdot Pr[\hat{f}_i^{(k)} \geq \tau ]^{T/2} &\quad& \textit{Simplify terms}\\ 
    &\lTilde \frac{T}{2} \cdot \left( 2e \right)^{T/2} \cdot \left(\frac{1}{8}\right)^{\frac{T}{2}} &\quad& \textit{Plug in upper bound for $Pr[\hat{f}_i^{(k)} \geq \tau ]$}\\
    &\lTilde \frac{T}{2} \cdot \left( 2e \right)^{T/2} \cdot e^{-\ln(8)\frac{T}{2}}  &\quad& \textit{Use trick that $e^{-\ln(8)} = \frac{1}{8}$}\\
    &\lTilde \frac{T}{2} \cdot \left( 2e \right)^{T/2} \cdot e^{-\Theta(T)}  &\quad& \textit{Rewrite $e^{-\ln(8)\frac{T}{2}}$ as $e^{-\Theta(T)}$}\\
    &\lTilde  e^{-\Theta(T)} &\quad& \textit{Omit further constant parts of term}\\
    &\lTilde e^{-\Theta(\log \log n)}  &\quad& \textit{Plug in value for $T$}\\
    &\lTilde \frac{1}{\log^C(n)}  &\quad &\textit{Rewrite term; plug in constant $C$ for $\Theta$}\\
    &\lTilde \frac{1}{\log^{100}(n)} &\quad &\textit{Plug in concrete value for C}
\end{flalign*}
Now that we have found an upper bound for $p_S(G_2)$, the probability that the bad scenario occurs, i.e., the algorithm outputs the estimate from the big sketch table for a rare element, we can upper bound the error $E_2$. 
By upper-bounding the probability for the good scenario $p_F(G_2)$ by one, we get:
\begin{flalign*}
E_2 &= \sum_{i=\frac{2A}{\tau}}^{n}  \E[f_i \cdot|f_i- \hat{f}_i|] \\ 
&\lTilde \sum_{i=\frac{2A}{\tau}}^{n} \left( p_S(G_2) \cdot  E_S + p_F(G_2) \cdot  E_F\right)  \\
&\lTilde\sum_{i=\frac{2A}{\tau}}^{n} \left( p_S(G_2) \cdot \frac{A^2}{iB} + p_F(G_2) \cdot \frac{A^2}{i^2} \right)&\quad & \textit{Plug in $E_S = \frac{A^2}{iB}$ and $E_F = \frac{A^2}{i^2}$} \\
&\lTilde \sum_{i=\frac{2A}{\tau}}^{n}  \left( \frac{A^2}{\log(n) ^{100}} \cdot \frac{1}{iB} +  \frac{A^2}{i^2} \right)&\quad & \textit{Plug in upper bound for $p_S(G_2)$ and upper bound $p_F(G_2)$ by one} \\
&\lTilde \frac{A^2}{B \cdot \log^{100}(n)} \cdot \sum_{i=\frac{2A}{\tau}}^{n} \frac{1}{i} + A^2 \cdot \sum_{i=\frac{2A}{\tau}}^{n} \frac{1}{i^2}  &\quad & \textit{Split sum and factor out constant parts}\\
&\lTilde \frac{A^2}{B \cdot \log^{100}(n)} \cdot \log(n)+  A^2 \cdot \int_{\frac{2A}{\tau}}^{n} \frac{1}{x^2} \, dx &\quad & \textit{Upper bound right sum by integral and left side by }\\
& & \quad & \textit{harmonic series $H(n) = \sum_{x=1}^n \frac{1}{x}\lTilde \log(n)$} \\
&\lTilde A^2 \cdot \left[\frac{1}{B \cdot \log^{99}(n)} + \left(- \frac{1}{n}+ \frac{\tau}{2A}  \right) \right] & \quad & \textit{Factor out$A^2$ and calculate integral} \\ 
&\lTilde A^2 \cdot \left[\frac{1}{B \cdot \log^{99}(n)} + \frac{\log \log n}{B} \right] &\quad & \textit{Plug in value for $\tau$, simplify, and omit constant parts}\\ 
& = \bigO \left(\frac{A^2\log\log n}{B} \right) &\quad & \textit{Only keep dominating part}
\end{flalign*}
\textbf{Case 3:} The last case to consider is the case of all elements $i$ such that
\[ \frac{A}{2\tau} \leq i \leq \frac{2A}{\tau} \xrightarrow{}  2\tau \geq f_i \geq \frac{\tau}{2} \]
In the worst case, the algorithm outputs zero for all these elements, since
\[\frac{A}{i} \geq \frac{\tau}{2} \gTilde \frac{A\log \log n}{2B}>\frac{A}{B}\]
Therefore, we can upper bound the error by assuming the worst-case scenario for all elements in group three:
\begin{flalign*}
E_3 &=  \E[f_i \cdot|f_i- \hat{f}_i|] \\ 
& \lTilde \sum_{i=\frac{A}{2\tau}}^{\frac{2A}{\tau}} \left( p_S(G_3) \cdot  E_S + p_F(G_3) \cdot  E_F\right)\\
& \lTilde \sum_{i=\frac{A}{2\tau}}^{\frac{2A}{\tau}} \frac{A^2}{i^2} \qquad & \textit{Assume worst case, i.e. $p_F(G_3) = 1$} \\
& \lTilde A^2 \cdot \int_{\frac{A}{2\tau}}^{\frac{2A}{\tau}} \frac{1}{x^2} \, dx \qquad & \textit{Upper bound sum by integral}\\
& \lTilde A^2 \left(-\frac{\tau}{2A} + \frac{2\tau}{A} \right) \qquad& \textit{Calculate the integral}\\
&\lTilde A^2 \left(-\frac{\log \log n}{2B} + \frac{2 \log \log n}{B} \right) \qquad &\textit{Plug in value for $\tau$}\\
&= \bigO \left( \frac{A^2 \log \log n}{B} \right) \\
\end{flalign*}
Finally, we can combine the three cases to obtain the total expected weighted error:
\begin{align*}
E_{total} &= \frac{1}{N} (E_1 + E_2 + E_3 ) = \frac{1}{N} \left[\bigO \left( \frac{A^2 \log B}{B} \right)+  \bigO \left( \frac{A^2  \log \log n}{B} \right) + \bigO \left( \frac{A^2 \log \log n}{B} \right) \right]\\
&= \frac{1}{N}\cdot \bigO \left( \frac{A^2 \log B}{B} \right)  = \frac{1}{N} \cdot  \bigO \left( \frac{N^2}{\log^2(n)} \frac{ \log B}{B} \right) \\
&= \bigO \left( \frac{ N}{\log^2(n)} \frac{ \log B}{B} \right) 
\qedhere
\end{align*}
since $\log B > \log \log n$. \\

This concludes the improved analysis of the weighted error for standard CountSketch++. It establishes an improved upper bound of $\bigO \left( \frac{N}{\log^2(n)}\frac{\log B}{B} \right)$ as stated in~\cref{thm:improved_error}.
\end{proof}

%% file: Chapters/appendix_large_sketch.tex
\section{Equalizing learned CountSketch++ without using predictions}\label{appendix:proof_large_sketch}
In the following, we provide the full proof of ~\cref{cor} as stated in Section~\cref{theory1}.
It follows the same structure as the proof of ~\cref{thm:improved_error} in~\cref{theory1}, with only minor changes to the variables. \\

We restate the corollary here for convenience:
\corLearnedSketch*

Intuitively, the algorithm partitions the elements into three groups (cf.~\cref{fig:element_groups}): Group one contains the heaviest elements. Group two comprises the rarest elements.  
Finally, group three holds all elements in between whose frequencies are close to the threshold $\tau$ used to determine whether an element is heavy or not during the filtering step.
We show that the algorithm outputs the estimate from the big sketch for the heaviest elements (i.e., group one) and outputs zero for the rarest elements (i.e., group two) with high probability. \\

To match the performance of a learned CountSketch++ using space $B$ with a standard CountSketch++ (i.e., without predictions), choose the following parameters:
\begin{enumerate} 
    \item  Threshold to output zero: $\tau = \frac{A \cdot \bigO(1)}{B}$
    \item  Total space budget: $B' = \bigO(B  \log B)$,
    \item  Width of the small sketch tables: $w = \bigO(B)$,
    \item  Width of the big sketch table: $w' = \bigO(B \log B)$.
\end{enumerate}

\begin{proof}
We are going to use the following theorem as stated and proved in~\citet{MIT}: 
\begin{theorem}\label{thm:MIT}
Let $\hat{f}_i$ be the estimate of the $i^{th}$ frequency given by a 3 x $\frac{B'}{3}$ CountSketch table. There exists a universal constant C such that the following inequality holds: 
\[\forall t \geq 3/B', \qquad Pr[|f_i-\hat{f}_i| \geq t] \leq C' \left( \frac{\log(tB')}{tB'}\right)^2\]
\end{theorem}
First, we decompose the total expected weighted error, {$\mathbf{E_{total}}$}, into three parts. Each part represents the error incurred by the respective group of elements, 
where group one contains heavy hitters, group two represents rare elements, and group three the elements in between (cf.~\cref{fig:element_groups}): 
\[E_{total} = \frac{1}{N} \sum_{i \in [n]} f_i \cdot |\hat{f_i}- f_i| = \frac{1}{N} (E_1 + E_2 + E_3) \]
For all groups of elements, let us decompose the expected weighted error $\mathbf{E_x, x \in \{1,2,3\}}$ they contribute to the final error $E_{total}$ into two parts $E_S$ and $E_F$. 
$\mathbf{E_S}$ describes the error obtained if we output the estimate in the sketch. $\mathbf{E_F}$ is the error incurred by outputting zero.
Weighting each part, $E_S$ and $E_F$, by the probability that these events occur (i.e., outputting zero vs. outputting the estimate from the big sketch table) leads to: 
\[E_x =  \sum_{i \in [G_x]}\left( p_S(G_x) \cdot  E_S + p_F(G_x) \cdot  E_F\right)\]
where $\mathbf{G_x, x \in \{1,2,3\}}$ is the group of elements under consideration, $\mathbf{p_S(G_x)}$ is the probability that the algorithm outputs the estimate from the big table for elements in group $G_x$, and $\mathbf{p_F(G_x)}$ is the probability that the algorithm outputs zero. 
Additionally, we know that for all groups: 
\[E_S = \E[f_i \cdot|f_i- \tilde{f}_i|] = \frac{A}{i} \cdot \frac{A}{B \log B} = \frac{A^2}{iB \log B}\]
\[E_F = \E[f_i \cdot|f_i- 0|] = \frac{1}{i} \cdot \frac{A}{i} = \frac{A}{i^2}\]
Since for all $i\leq B \log B$ it holds that $E_S = \frac{A}{i} \cdot \frac{A}{B \log B} \leq  \frac{A^2}{i^2} = E_F$, it is desirable to output the estimate from the big CountSketch table $\tilde{f}_i$ for these elements. 
For all other elements, it follows that it is preferable to output zero.  \\ 

\textbf{Case 1:} This case deals with all elements $i$, such that 
\[i\leq\frac{A}{2\tau} = \frac{B}{2} \xrightarrow{} f_i= \frac{A}{i}\geq 2\tau = \frac{2 A }{B}\]
For these elements, it is desirable that the algorithm outputs the estimate from the big sketch table. 
Accordingly, the bad scenario is that the algorithm labels these frequent elements as rare and outputs zero, since 
\[\forall i \leq \frac{A}{2 \tau}: \qquad f_i = \frac{A}{i} \geq \frac{2 A}{B} > \frac{A}{B \log B}\] 
To upper bound the error $E_1$, we aim to minimize the upper bound of the probability that the algorithm outputs zero, $p_F$. 
Let $\hat{f}_i^{(k)}$ be the frequency estimate of the $i^{th}$ element given by the $k^{th}$ small CountSketch table.
In the following, $\lTilde$ denotes inequality up to a constant factor. 
The probability that a single estimate from a small table $\hat{f}_i^{(k)}$ is smaller than $\tau$ is:
\begin{alignat*}{2}
\Pr[\hat{f}_i^{(k)} \leq \tau ] &= \Pr\left[-\hat{f}_i^{(k)} \geq -\tau\right] &\quad& \textit{Multiply both sides by -1}\\
&= \Pr\left[f_i - \hat{f}_i^{(k)} \geq f_i - \tau\right] &\quad& \textit{Add $f_i$ to both sides} \\
&\leq \Pr\left[|f_i - \hat{f}_i^{(k)}| \geq f_i - \tau\right] &\quad& \textit{Take the absolute value of the left side}  \\
&\leq \Pr\left[|f_i - \hat{f}_i^{(k)}| \geq 2 \tau - \tau\right] &\quad& \textit{Lower bound $f_i$ by $2\tau$, the minimum frequency in group 1} \\
&\leq \Pr\left[|f_i - \hat{f}_i^{(k)}| \geq \tau\right] \\
\end{alignat*}

Given that term, we can now apply~\cref{thm:MIT}:
\begin{alignat*}{2}
\Pr\left[|f_i - \hat{f}_i^{(k)}| \geq \tau\right]
&\leq C'\left(\frac{\log(\tau B')}{\tau B'}\right)^2 &\quad& \textit{Apply \cref{thm:MIT}} \\
&\leq C'\left(\frac{\log(\frac{C  A }{B} \cdot \frac{B\log B}{2T})}{\frac{C A }{B} \cdot \frac{B\log B}{2T}}\right)^2 &\quad& \textit{Plug in values: $\tau= \frac{A \cdot \bigO(1)}{B}$ and $B'= \frac{B \log B}{2T}$} \\
&\leq C'\left(\frac{\log(\frac{C  A }{B} \cdot \frac{B\log B}{2C''\log\log n})}{\frac{C A }{B} \cdot \frac{B\log B}{2C''\log\log n}}\right)^2 &\quad& \textit{Plug in value for $T, T= \bigO(\log \log n)$} \\
&\leq C'\left(\frac{2C'' \log\left(\frac{AC}{2C''}\right)}{AC}\right)^2 &\quad& \textit{Reduce terms where possible} \\
&\leq 4C'C''^2\left(\frac{\log\left(\frac{AC}{2C''}\right)}{AC}\right)^2 &\quad& \textit{Move 2C'' out of parentheses} \\
&\lTilde \left(\frac{\log(AC)}{AC}\right)^2 \lTilde \frac{1}{8}&\quad& \textit{Upper bound by $\frac{1}{8}$} \\
&\lTilde e^{-\ln(8) } &\quad& \textit{Use trick that $e$ and $\ln$ equalize each other} 
\end{alignat*}

Let $m$ be the number of small tables with estimates smaller than $\tau$. If the algorithm outputs zero, at least half the estimates from these $T$ small tables are smaller than the threshold $\tau$, therefore,  $m \geq \frac{T}{2}$.
There are $\binom{T}{m}$ different combinations to choose $m$ from the $T$ estimates to be smaller than the threshold $\tau$.  
Given that the estimates are independent, and the probability that a single estimate from a small table is smaller than $\tau$ is upper bounded by $e^{-\ln(8)}$ as stated above, we get: 
\begin{alignat*}{2} 
    p_F(G_1) &= \sum_{m=T/2}^T \binom{T}{m} \cdot Pr[\hat{f}_i^{(k)} \leq \tau ]^m \cdot Pr[\hat{f}_i^{(k)} > \tau ]^{T-m} \\
    &\leq \sum_{m=T/2}^T \binom{T}{m} \cdot Pr[\hat{f}_i^{(k)} \leq \tau ]^m &\quad& \textit{Omit $Pr[\hat{f}_i^{(k)} > \tau ]^{T-m}$} \\ 
    & \leq \frac{T}{2} \cdot \binom{T}{T/2} \cdot Pr[\hat{f}_i^{(k)} \leq \tau ]^{T/2} &\quad& \textit{Upper bound the single addends $\left(\textit{max. for } m= \frac{T}{2}\right)$} \\ 
    & \leq \frac{T}{2} \cdot \left( \frac{2eT}{T} \right)^{T/2} \cdot Pr[\hat{f}_i^{(k)} \leq \tau ]^{T/2} &\quad& \textit{Upper bound binom. coefficient by $\binom{n}{k} \leq \left(\frac{e n}{k}\right)^k$}\\ 
    & \leq \frac{T}{2} \cdot \left( 2e \right)^{T/2} \cdot Pr[\hat{f}_i^{(k)} \leq \tau ]^{T/2} &\quad& \textit{Reduce terms}\\ 
    &\lTilde \frac{T}{2} \cdot \left( 2e \right)^{T/2} \cdot e^{-\ln(8)\frac{T}{2}} &\quad& \textit{Plug in upper bound for $Pr[\hat{f}_i^{(k)} \leq \tau ]$}\\
    &\lTilde \frac{T}{2} \cdot \left( 2e \right)^{T/2} \cdot e^{-\Theta(T)} &\quad& \textit{Rewrite $e^{-\ln(8)\frac{T}{2}}$ as $e^{-\Theta(T)}$}\\
    &\lTilde e^{-\Theta(T)}  &\quad& \textit{Omit further constant parts of term} \\
    &\lTilde e^{-\Theta(\log \log n)}  &\quad& \textit{Plug in value for $T$} \\
    &\lTilde \frac{1}{\log^C(n)} &\quad &\textit{Rewrite term; plug in constant $C$ for $\Theta$} \\
    & \lTilde \frac{1}{\log^{100}(n)} &\quad &\textit{Plug in concrete value for C}
\end{alignat*}

Plugging in the obtained upper bound for $p_F(G_1)$, the probability that the bad scenario occurs, i.e., the algorithm outputs zero, and upper-bounding the probability for the good scenario $p_S(G_1)$ by one, we get: 
\begin{alignat*}{2}
E_1 &= \sum_{i=1}^{\frac{A}{2\tau}} \E[f_i \cdot|f_i- \hat{f}_i|] \\
&= \sum_{i=1}^{\frac{A}{2\tau}} \left( p_F(G_1) \cdot  E_F +p_S(G_1) \cdot  E_S\right) \\
&\lTilde \sum_{i=1}^{\frac{A}{2\tau}} \left(p_F(G_1) \cdot \frac{A^2}{i^2} + p_S(G_1) \cdot \frac{A^2}{iB\log B}  \right)  &\quad & \textit{Plug in $E_S = \frac{A^2}{iB\log B}$ and $E_F = \frac{A^2}{i^2}$} \\
&\lTilde \sum_{i=1}^{\frac{A}{2\tau}} \left( \frac{1}{\log^{100}(n)} \cdot \frac{A^2}{i^2} + \frac{A^2}{iB\log B} \right)  &\quad & \textit{Plug in obtained upper bound for $p_F(G_1)$ } \\
&& \quad & \textit{and upper bound $p_S(G_1)$ by 1} \\ 
&\lTilde \frac{A^2}{\log^{100}(n)} \sum_{i=1}^{\frac{A}{2\tau}} \frac{1}{i^2} + \frac{A^2}{B\log B} \sum_{i=1}^{\frac{A}{2\tau}} \frac{1}{i}  &\quad & \textit{Move constants out of sum}\\
&\lTilde \frac{A^2}{\log^{100}(n)} \int_1^{\frac{A}{2\tau}} \frac{1}{x^2} \, dx + \frac{A^2}{B\log B}  \cdot \log\left(\frac{A}{2\tau} \right) &\quad & \textit{Upper bound left sum by integral and right sum}\\
& & \quad & \textit{by harmonic series $H(n) = \sum_{x=1}^n \frac{1}{x}\lTilde \log(n)$} \\
&\lTilde \frac{A^2}{\log^{100}(n)} \cdot \left(1- {\frac{2\tau}{A}}\right) + \frac{A^2}{B\log B} \cdot \log\left(\frac{A}{2\tau} \right)&\quad & \textit{Calculate integral}\\
&\lTilde \frac{A^2}{\log^{100}(n)} \cdot \left( 1 - {\frac{2}{B}}\right) + \frac{A^2}{B\log B} \cdot \log\left(\frac{B}{2} \right) &\quad & \textit{Plug in value for $\tau$}\\
&\lTilde A^2 \left(\frac{B-2}{\log^{100}(n) B} + \frac{\log (B/2)}{B\log B}\right) &\quad & \textit{Pull out $A^2$ and pull apart logarithmic term}\\
&=\bigO\left(\frac{A^2 \log (B/2)}{B\log B}\right) =\bigO\left(\frac{A^2}{B}\right) &\quad & \textit{Only keep dominating part and reduce}
\end{alignat*}

\textbf{Case 2:} This case deals with all elements $i$, such that
\[i\geq \frac{2A}{\tau} = 2B \xrightarrow{} f_i= \frac{A}{i}\leq \frac{\tau}{2} = \frac{A}{2B}\]
Note that not for all elements in this group, the best and worst cases are the same: For some elements $i$ such that $2B \leq i \leq B \log B$, it is indeed still preferable to output the estimate from the big sketch table.
However, for most elements in group two, the best and worst case scenarios are just inverse to the first case:
It is desirable that the algorithm outputs zero, while the bad scenario is that the algorithm outputs the estimate from the big sketch table. \\ 

Following the same reasoning as in the first case, we can upper bound the error $E_2$ by upper bounding the probability that the algorithm outputs the estimate from the big sketch table, $p_S$.
To do so, let us first consider the probability that a single estimate from the small tables is larger than $\tau$:
\begin{alignat*}{2}
\Pr[\hat{f}_i^{(k)} \geq \tau ] &= \Pr\left[\hat{f}_i^{(k)}-f_i \geq  \tau - f_i\right]  &\quad& \textit{Subtract $f_i$ from both sides} \\
&\leq \Pr\left[|\hat{f}_i^{(k)}-f_i| \geq  \tau -f_i \right]  &\quad&  \textit{Take the absolute value of the left side}  \\
&\leq \Pr\left[|\hat{f}_i^{(k)}-f_i | \geq \tau - \frac{\tau}{2}\right] &\quad& \textit{Upper bound $f_i$ by $\frac{\tau}{2}$, the max. frequency in group 2} \\ 
&\leq \Pr\left[| \hat{f}_i^{(k)}-f_i | \geq \frac{\tau}{2}\right] \\
\end{alignat*}

Like in the first case, we can now apply \cref{thm:MIT}:
\begin{alignat*}{2}
\Pr\left[|f_i - \hat{f}_i^{(k)}| \geq \frac{\tau}{2}\right] &\leq C'\left(\frac{\log(\frac{\tau}{2}B')}{\frac{\tau}{2}B'}\right)^2 &\quad& \textit{Apply \cref{thm:MIT}} \\
&\leq C'\left(\frac{\log(\frac{CA}{2B} \cdot \frac{B \log B}{2T})}{ \frac{CA }{2B} \cdot \frac{B \log B}{2T}}\right)^2 &\quad& \textit{Plug in values: $\tau= \frac{A\bigO(1)}{B}$ and $B'= \frac{B \log B}{2T}$} \\
&\leq C'\left(\frac{\log(\frac{CA}{2B} \cdot \frac{B \log B}{2C''\log\log n})}{ \frac{CA }{2B} \cdot \frac{B \log B}{2C''\log\log n}}\right)^2 &\quad& \textit{Plug in value for $T, T= \bigO(\log \log n)$} \\
&\leq C'\left(\frac{4C'' \log\left(\frac{AC}{4C''}\right)}{AC}\right)^2  &\quad& \textit{Reduce terms where possible} \\
&\leq 16C'C''^2\left(\frac{\log\left(\frac{AC}{4C''}\right)}{AC}\right)^2 &\quad& \textit{Move constants out of parentheses} \\
&\lTilde \left(\frac{\log(AC)}{AC}\right)^2 \lTilde \frac{1}{8} &\quad& \textit{Omit irrelevant constants and upper bound by $\frac{1}{8}$} \\
&\lTilde e^{-\ln(8)}  &\quad&  \textit{Use trick that $e$ and $\ln$ equalize each other} 
\end{alignat*}

We can now determine the probability that the algorithm outputs the estimate from the big table, $p_S(G_2)$. The steps follow the same logic as in case one: 
\begin{alignat*}{2} 
    p_S(G_2) &= \sum_{m=T/2}^T \binom{T}{m} \cdot Pr[\hat{f}_i^{(k)} \geq \tau ]^m \cdot Pr[\hat{f}_i^{(k)} < \tau ]^{T-m} \\
    &\leq \sum_{m=T/2}^T \binom{T}{m} \cdot Pr[\hat{f}_i^{(k)} \geq \tau ]^m &\quad& \textit{Omit $Pr[\hat{f}_i^{(k)} < \tau ]^{T-m}$} \\
    & \leq \frac{T}{2} \cdot \binom{T}{T/2} \cdot Pr[\hat{f}_i^{(k)} \geq \tau ]^{T/2} &\quad& \textit{Upper bound the single addends $\left(\textit{max. for } m= \frac{T}{2}\right)$} \\ 
    & \leq \frac{T}{2} \cdot \left( \frac{2eT}{T} \right)^{T/2} \cdot Pr[\hat{f}_i^{(k)} \geq \tau ]^{T/2}  &\quad& \textit{Upper bound binom. coefficient by $\binom{n}{k} \leq \left(\frac{e n}{k}\right)^k$}\\
    &\leq \frac{T}{2} \cdot \left( 2e \right)^{T/2} \cdot Pr[\hat{f}_i^{(k)} \geq \tau ]^{T/2} &\quad& \textit{Reduce terms}\\
    &\lTilde \frac{T}{2} \cdot \left( 2e \right)^{T/2} \cdot e^{-\ln(8)\frac{T}{2}} &\quad& \textit{Plug in upper bound for $Pr[\hat{f}_i^{(k)} \geq \tau ]$}\\
    &\lTilde \frac{T}{2} \cdot \left( 2e \right)^{T/2} \cdot e^{-\Theta(T)}  &\quad& \textit{Rewrite $e^{-\ln(8)\frac{T}{2}}$ as $e^{-\Theta(T)}$}\\
    &\lTilde  e^{-\Theta(T)} &\quad& \textit{Omit further constant parts of term}\\
    &\lTilde e^{-\Theta(\log \log n)}  &\quad& \textit{Plug in value for $T$}\\
    &\lTilde \frac{1}{\log^C(n)}  &\quad &\textit{Rewrite term; plug in constant $C$ for $\Theta$}\\
    &\lTilde \frac{1}{\log^{100}(n)} &\quad &\textit{Plug in  concrete value for C}
\end{alignat*}

Hence: 
\begin{alignat*}{2}
E_2 &= \sum_{i=\frac{2A}{\tau}}^{n}  \E[f_i \cdot|f_i- \hat{f}_i|] \\ 
&\lTilde \sum_{i=\frac{2A}{\tau}}^{n} \left( p_S(G_2) \cdot  E_S + p_F(G_2) \cdot  E_F\right)  \\
&\lTilde\sum_{i=\frac{2A}{\tau}}^{n} \left( p_S(G_2) \cdot \frac{A^2}{iB \log B} + p_F(G_2) \cdot \frac{A^2}{i^2} \right)&\quad & \textit{Plug in $E_S = \frac{A^2}{iB \log B}$ and $E_F = \frac{A^2}{i^2}$} \\
&\lTilde \sum_{i=\frac{2A}{\tau}}^{n}  \left( \frac{A^2}{\log(n) ^{100}} \cdot \frac{1}{iB\log B} +  \frac{A^2}{i^2} \right)&\quad & \textit{Plug in upper bound for $p_S(G_2)$ } \\
&& \quad & \textit {and upper bound $p_F(G_2)$ by 1} \\
&\lTilde \frac{A^2}{B \log B \cdot \log^{100}(n)} \cdot \sum_{i=\frac{2A}{\tau}}^{n} \frac{1}{i} + A^2 \cdot \sum_{i=\frac{2A}{\tau}}^{n} \frac{1}{i^2}  &\quad & \textit{Split sum and pull out constant parts}\\
&\lTilde \frac{A^2}{B \log B\cdot \log^{100}(n)} \cdot \log(n)+  A^2 \cdot \int_{\frac{2A}{\tau}}^{n} \frac{1}{x^2} \, dx &\quad & \textit{Upper bound right sum by integral and left side by }\\
& & \quad & \textit{harmonic series $H(n) = \sum_{x=1}^n \frac{1}{x}\lTilde \log(n)$} \\
&\lTilde A^2 \cdot \left[\frac{1}{B \log B\cdot \log^{99}(n)} + \left(- \frac{1}{n}+ \frac{\tau}{2A}  \right) \right] & \quad & \textit{Pull out $A^2$ and calculate integral} \\ 
&\lTilde A^2 \cdot \left[\frac{1}{B\log B \cdot \log^{99}(n)} + \frac{1}{2B} \right] &\quad & \textit{Plug in value for $\tau$ and reduce}\\ 
& = \bigO \left(\frac{A^2}{B} \right) &\quad & \textit{Only keep dominating part}
\end{alignat*}

\textbf{Case 3:} The last case to consider is the case of all elements $i$ such that
\[ \frac{A}{2\tau} \leq i \leq \frac{2A}{\tau} \xrightarrow{}  2\tau \geq f_i \geq \frac{\tau}{2} \]
In the worst case, the algorithm outputs zero for all these elements, since
\[\frac{A}{i} \geq \frac{\tau}{2} \gTilde \frac{A}{2B}>\frac{A}{B \log B}\]
Therefore, we can upper bound the error by assuming the worst-case scenario for all elements:
\begin{alignat*}{2}
E_3 &=  \E[f_i \cdot|f_i- \hat{f}_i|] \\ 
& \lTilde \sum_{i=\frac{A}{2\tau}}^{\frac{2A}{\tau}} \left( p_S(G_3) \cdot  E_S + p_F(G_3) \cdot  E_F\right)\\
& \lTilde \sum_{i=\frac{A}{2\tau}}^{\frac{2A}{\tau}} \frac{A^2}{i^2} &\qquad& \textit{Assume worst case, i.e. $p_F(G_3) = 1$} \\
& \lTilde A^2 \cdot \int_{\frac{A}{2\tau}}^{\frac{2A}{\tau}} \frac{1}{x^2} \, dx& \qquad& \textit{Upper bound sum by integral}\\
& \lTilde A^2 \left(-\frac{\tau}{2A} + \frac{2\tau}{A} \right) & \qquad & \textit{Calculate the integral}\\
&\lTilde A^2 \left(-\frac{1}{2B } + \frac{2}{B } \right) & \qquad & \textit{Plug in value for $\tau$}\\
&= \bigO \left( \frac{A^2}{B } \right) \\
\end{alignat*}

Finally, we can combine the three cases to obtain the total expected weighted error:
\begin{align*}
E_{total} &= \frac{1}{N} (E_1 + E_2 + E_3 ) = \frac{1}{N} \left[\bigO \left( \frac{A^2 }{B} \right)+  \bigO \left( \frac{A^2}{B} \right) + \bigO \left( \frac{A^2}{B } \right) \right]\\
&= \frac{1}{N}\cdot \bigO \left( \frac{A^2 }{B} \right)  = \frac{1}{N} \cdot  \bigO \left( \frac{N^2}{\log^2(n)} \frac{1}{B} \right) \\
&= \bigO \left( \frac{ N}{\log^2(n)} \frac{1}{B} \right) 
\end{align*} 
\qedhere
\end{proof}
This concludes the proof that, indeed, a standard CountSketch++ using a space budget of $B' = \bigO(B \log B)$ achieves the same weighted error as its learned counterpart with space budget $B$.

%% file: Chapters/appendix_trivial_alternative.tex
\section{A Trivial Alternative to learned frequency estimation}\label{appendix:trivial_alternative}
First, let us recap some insights from earlier sections to intuitively understand why the Trivial Alternative asymptotically performs as well as learning-augmented algorithms:
Since the learned CountSketch++ algorithm only uses the predictions to store the top-$\frac{B}{2}$ elements exactly, the error solely comes from the remaining elements $i$, 
such that $i \in [n] \setminus [\frac{B}{2}] \Rightarrow i \geq \frac{B}{2} \Rightarrow f_i \leq \frac{2A}{B}$. \\

\begin{figure}[h]
    \centering
    \begin{tikzpicture}[scale=1.2]
    \fill[blue!20, opacity=0.5] (0,0) -- (2.8,0) -- (2.8,5) -- (0,5) -- cycle;

    \draw[->] (0,0) -- (7,0) node[right] {$i$};
    \draw[->] (0,0) -- (0,5) node[above] {};
    \draw (0.5,0.1) -- (0.5,-0.1) node[below] {$\frac{A}{2\tau}$};
    \draw (2,0.1) -- (2,-0.1) node[below] {$\frac{2A}{\tau}$};
    \draw (1.1,0.1) -- (1.1,-0.1) node[below] {$\frac{A}{\tau}$};
    \draw (2.8,0.1) -- (2.8,-0.1) node[below] {$\frac{B}{2}$};

    \draw (-0.1,0.9) -- (0.1,0.9);
    \node at (-0.25,0.9) {$\tau$};
    \draw[dashed] (0.5,0) -- (0.5,5);
    \draw[dashed] (2,0) -- (2,5);
    \draw[dashed] (1.1,0) -- (1.1,0.9);
    \draw (2.8,0) -- (2.8,5);

    \draw[dashed] (7,0.9) -- (0,0.9);
    \draw[thick] plot[domain=0.2:7, samples=100] (\x, {1/\x});

    \node at (4.2,2.5) {Exact Counts for };
    \node at (4.2,2.1) { top-$\frac{B}{2}$ Elements};
    \draw[->, thick] (3.1,2.25) -- (1.7,2.25);

    \node[rotate=90] at (-0.5,2.5) {$f_i$};
    \end{tikzpicture}
    \caption[Intuition for Trivial Mechanism]{Given that the top-$\frac{B}{2}$ elements are known and tracked exactly, the algorithm should output zero for all remaining elements, since their frequencies are (asymptotically) smaller than $\tau$. }
    \label{fig:triv_alt}
\end{figure}
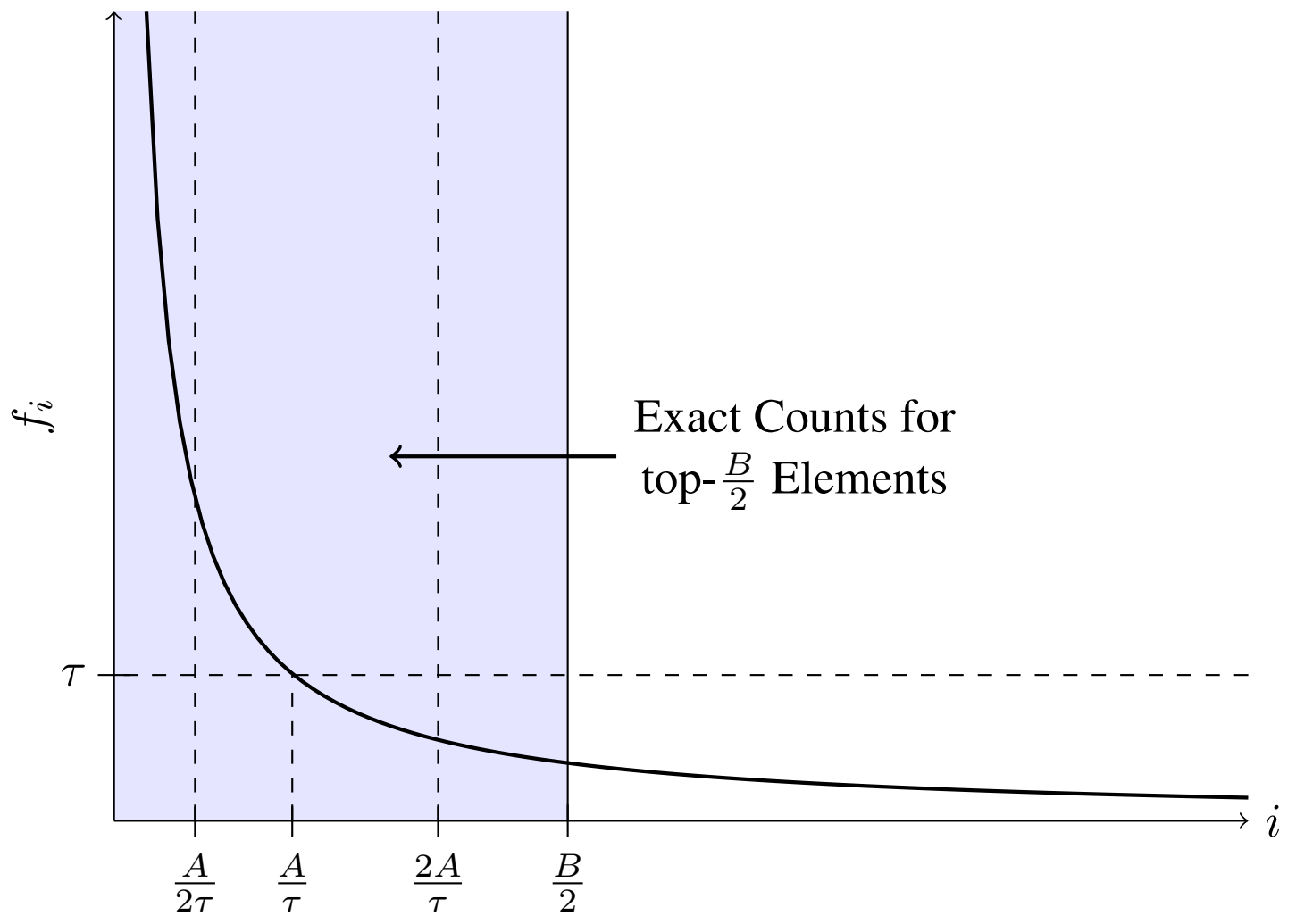

If CS++ works well, the estimate from the small tables should be close to the true element frequencies. 
More specifically, the estimate should be asymptotically smaller than $\tau = \frac{A\bigO(\log \log n)}{B}$ for all of these remaining elements that are not predicted to be in the top-$\frac{B}{2}$. This means that if the algorithm works as desired, it should output zero for all elements that are not tracked with exact counters in a separate array (cf.~\cref{fig:triv_alt}). \\

Instead of setting up a data structure for the remaining elements, one could therefore take a shortcut: Only track the top-$\frac{B}{2}$ elements exactly, and return zero for all other elements.
This approach leads to the same asymptotic error as the learned CountSketch++ algorithm while using half the space: 

\begin{proof}
\begin{alignat*}{2}
    E_{total} &= \frac{1}{N} \left( \sum_{i=1}^{\lfloor \frac{B}{2} \rfloor} \E[f_i \cdot |\hat{f}_i - f_i|] + \sum_{i=\lfloor \frac{B}{2}\rfloor +1}^{n} \E[f_i \cdot |\hat{f}_i - 0|] \right) \\
    &= \frac{1}{N} \left(0 + \sum_{i=\lfloor \frac{B}{2} \rfloor +1}^{n} \E[f_i \cdot |\hat{f}_i - 0|]\right) & \quad & \textit{Error for top-$\frac{B}{2}$ elements is zero } \\[-6pt]
    && \quad & \textit{due to exact counts}\\
    & \lTilde \frac{1}{N}\sum_{i=\lfloor \frac{B}{2} \rfloor +1}^{n} \frac{A^2}{i^2} & \quad & \textit{Error when outputting zero for all remaining} \\[-6pt]
    && \quad & \textit{elements is $\frac{A^2}{i^2}$ per element} \\[-6pt]
    & \lTilde \frac{A^2}{N} \int_{\lfloor \frac{B}{2} \rfloor +1}^{n} \frac{1}{x^2} \, dx  \lTilde \frac{N^2}{\log(n)^2 N} \left(-\frac{1}{n} + \frac{2}{B} \right) & \quad & \textit{Upper bound the sum by the integral } \\
    & = \bigO \left( \frac{N}{\log^2(n)} \cdot \frac{1}{B} \right)
\end{alignat*}
\end{proof}

%% file: Chapters/appendix_experiments.tex
\section{Supplementary Experimental Results}\label{appendix:experiments}

\subsection{Synthetic Zipfian Data}
\vspace{20px}
\begin{figure}[H]
    \centering
    \includegraphics[height=0.8\textheight, width=\textwidth]{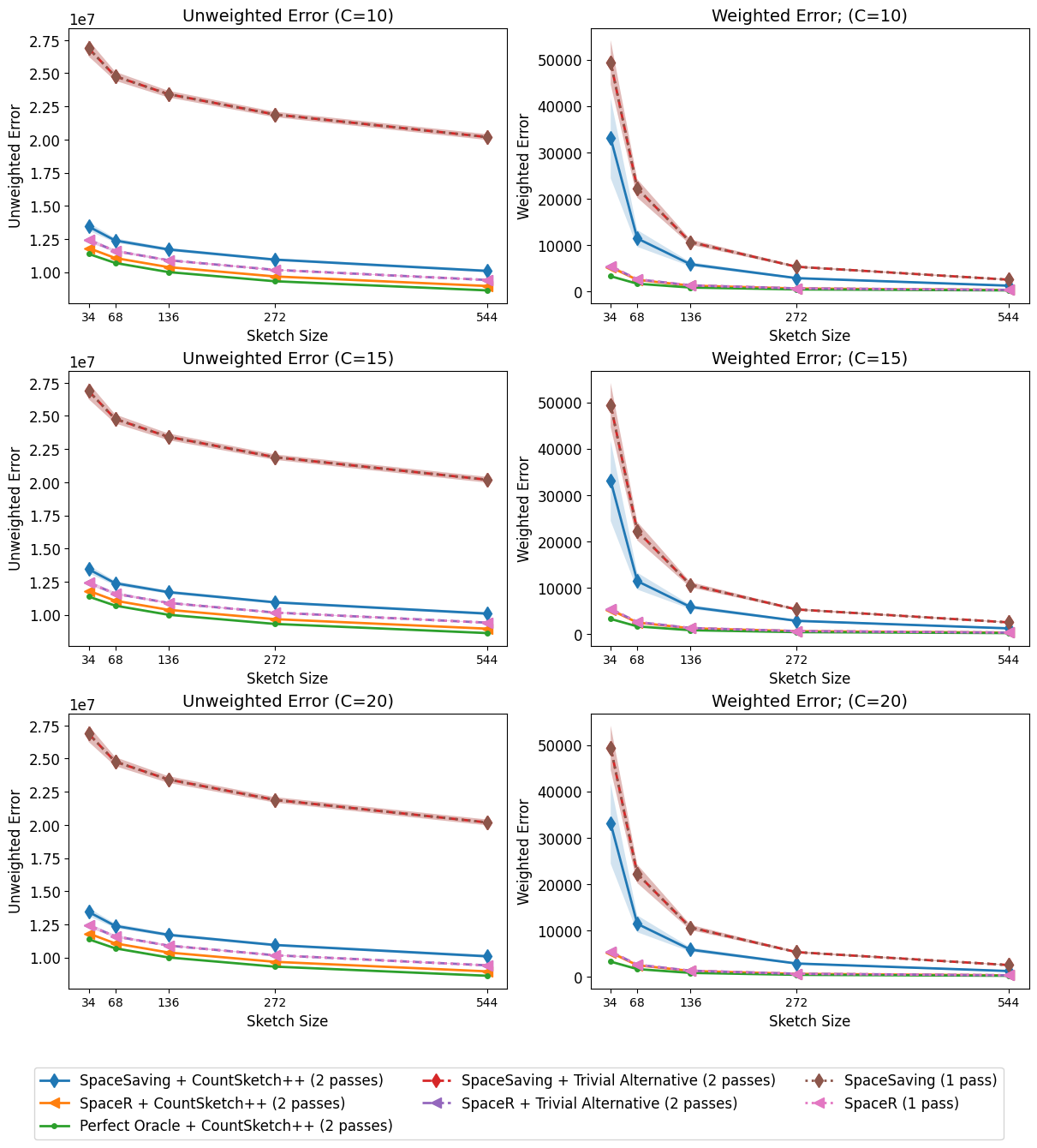}
    \caption[Results Experiment 2: One-Pass SpaceR vs. Two-Pass Learned
Frameworks]{Comparison of weighted and unweighted error between the best-performing approaches over one and two passes for the synthetic dataset for all tested values of the threshold parameter $C$.}
\end{figure}

\subsection{CAIDA Dataset}
\vspace{20px}
\begin{figure}[H]
    \centering
    \includegraphics[height=0.8\textheight, width=\textwidth]{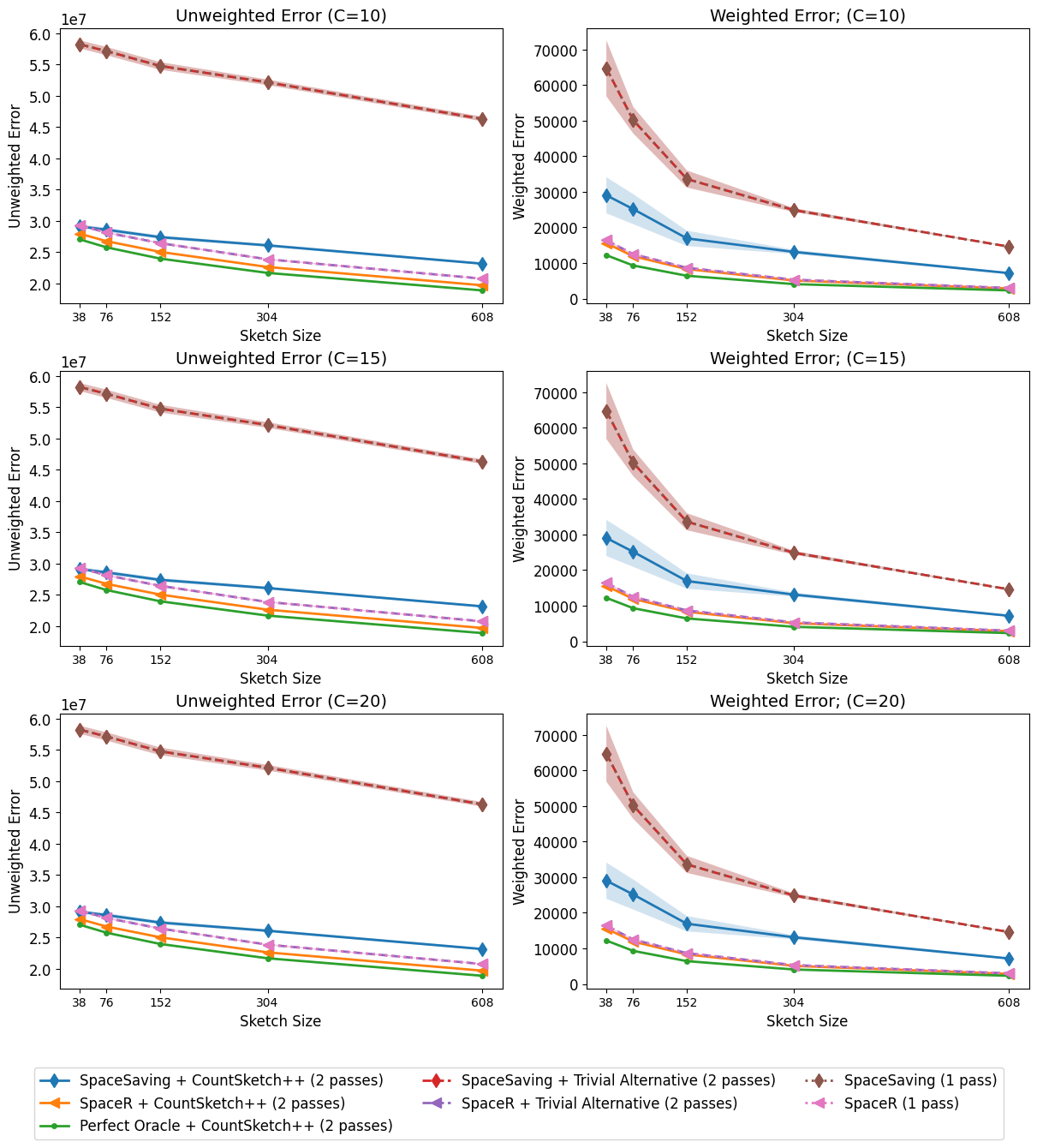}
   \caption[Results Experiment 2: One-Pass SpaceR vs. Two-Pass Learned
Frameworks]{Comparison of weighted and unweighted error between the best-performing approaches over one and two passes for the CAIDA dataset for all tested values of the threshold parameter $C$.}
\end{figure}

%% file: checklist.tex
\section*{NeurIPS Paper Checklist}

\begin{enumerate}

\item {\bf Claims}
    \item[] Question: Do the main claims made in the abstract and introduction accurately reflect the paper's contributions and scope?
    \item[] Answer: \answerYes{} 
    \item[] Justification: We provide both theoretical and experimental results in the main section of the paper. 

\item {\bf Limitations}
    \item[] Question: Does the paper discuss the limitations of the work performed by the authors?
    \item[] Answer: \answerYes{} 
    \item[] Justification: See \cref{conclusion} for the limitations discussed, e.g., theoretical guarantees for SpaceR.

\item {\bf Theory assumptions and proofs}
    \item[] Question: For each theoretical result, does the paper provide the full set of assumptions and a complete (and correct) proof?
    \item[] Answer: \answerYes{} 
    \item[] Justification: Proofs for the theoretical results are provided in
    \cref{appendix:improved_analysis}, \cref{appendix:proof_large_sketch}, and \cref{appendix:trivial_alternative}.

    \item {\bf Experimental result reproducibility}
    \item[] Question: Does the paper fully disclose all the information needed to reproduce the main experimental results of the paper to the extent that it affects the main claims and/or conclusions of the paper (regardless of whether the code and data are provided or not)?
    \item[] Answer: \answerYes{} 
    \item[] Justification: See \cref{experiments} for details on experiment setup and datasets. 

\item {\bf Open access to data and code}
    \item[] Question: Does the paper provide open access to the data and code, with sufficient instructions to faithfully reproduce the main experimental results, as described in supplemental material?
    \item[] Answer: \answerYes{} 
    \item[] Justification: We provide the code in the supplementary materials and note that for the CAIDA dataset, we are not allowed to distribute the data ourselves, but provide instructions on how to obtain it in the Readme.

\item {\bf Experimental setting/details}
    \item[] Question: Does the paper specify all the training and test details (e.g., data splits, hyperparameters, how they were chosen, type of optimizer) necessary to understand the results?
    \item[] Answer: \answerYes{} 
    \item[] Justification: See \cref{experiments} for the necessary details. 

\item {\bf Experiment statistical significance}
    \item[] Question: Does the paper report error bars suitably and correctly defined or other appropriate information about the statistical significance of the experiments?
    \item[] Answer: \answerYes{} 
    \item[] Justification: As described in \cref{experiments}, we run each experiment five times with different seeds and indicate the results in the plots accordingly. 

\item {\bf Experiments compute resources}
    \item[] Question: For each experiment, does the paper provide sufficient information on the computer resources (type of compute workers, memory, time of execution) needed to reproduce the experiments?
    \item[] Answer: \answerYes{} 
    \item[] Justification: We provided all necessary information in \cref{experiments} and deem hardware constraints beyond the details there not relevant for reproducing the (comparably lightweight) experiments.  
    
\item {\bf Code of ethics}
    \item[] Question: Does the research conducted in the paper conform, in every respect, with the NeurIPS Code of Ethics \url{https://neurips.cc/public/EthicsGuidelines}?
    \item[] Answer: \answerYes{} 
    \item[] Justification: There was no deviation from the NeurIPS Code of Ethics. 

\item {\bf Broader impacts}
    \item[] Question: Does the paper discuss both potential positive societal impacts and negative societal impacts of the work performed?
    \item[] Answer: \answerNA{} 
    \item[] Justification: The work operates within an established field of frequency estimation. In our view, there is no societal impact that needs to be addressed in the context of our contribution. 
    
\item {\bf Safeguards}
    \item[] Question: Does the paper describe safeguards that have been put in place for responsible release of data or models that have a high risk for misuse (e.g., pre-trained language models, image generators, or scraped datasets)?
    \item[] Answer: \answerNA{} 
    \item[] Justification: Paper poses no such risks

\item {\bf Licenses for existing assets}
    \item[] Question: Are the creators or original owners of assets (e.g., code, data, models), used in the paper, properly credited and are the license and terms of use explicitly mentioned and properly respected?
    \item[] Answer: \answerYes{} 
    \item[] Justification: The paper contains all necessary credits and licenses. 

\item {\bf New assets}
    \item[] Question: Are new assets introduced in the paper well documented and is the documentation provided alongside the assets?
    \item[] Answer: \answerNA{} 
    \item[] Justification: No new assets added. 

\item {\bf Crowdsourcing and research with human subjects}
    \item[] Question: For crowdsourcing experiments and research with human subjects, does the paper include the full text of instructions given to participants and screenshots, if applicable, as well as details about compensation (if any)? 
    \item[] Answer: \answerNA{} 
    \item[] Justification: No crowdsourcing or human subjects were involved in the experiments of this paper. 

\item {\bf Institutional review board (IRB) approvals or equivalent for research with human subjects}
    \item[] Question: Does the paper describe potential risks incurred by study participants, whether such risks were disclosed to the subjects, and whether Institutional Review Board (IRB) approvals (or an equivalent approval/review based on the requirements of your country or institution) were obtained?
    \item[] Answer: \answerNA{} 
    \item[] Justification: No crowdsourcing or human subjects were involved in the experiments of this paper. 

\item {\bf Declaration of LLM usage}
    \item[] Question: Does the paper describe the usage of LLMs if it is an important, original, or non-standard component of the core methods in this research? Note that if the LLM is used only for writing, editing, or formatting purposes and does \emph{not} impact the core methodology, scientific rigor, or originality of the research, declaration is not required.
    \item[] Answer: \answerNA{} 
    \item[] Justification: LLMs were only used for formatting and editing. 

\end{enumerate}